\documentclass[aps,prd,reprint,superscriptaddress,nofootinbib,amsfonts,amssymb,amsmath]{revtex4-1}

\usepackage[utf8x]{inputenc}
\usepackage{dsfont}
\usepackage{graphicx}
\usepackage{bm}
\usepackage{xcolor}
\usepackage{slashed}
\usepackage{url}
\usepackage{booktabs}
\usepackage{slashed}
\usepackage{microtype}

\allowdisplaybreaks

\graphicspath{{./figures/}}

\usepackage{hyperref}
\hypersetup{
colorlinks=true,
citecolor=blue,
citebordercolor=red,
linktoc=all,
linkcolor=blue,
urlcolor=blue
}

\def\s0#1#2{\mbox{\small{$ \frac{#1}{#2} $}}}
\def\0#1#2{\frac{#1}{#2}}

\newcommand{\qqquad}{\qquad \qquad}
\newcommand{\qqqquad}{\qquad \qquad \qquad}

\def\CI{{\mathcal I}}

\def\CO{{\mathcal O}}

\newcommand{\Tr}{\text{Tr}}

\newcommand{\tr}{\text{tr}}
\newcommand{\tinytr}{\text{\tiny tr}}
\newcommand{\TT}{\text{tt}}
\newcommand{\tinyTT}{\text{\tiny tt}}
\newcommand{\Flow}{\text{Flow}}
\newcommand{\gauge}{\text{\tiny gauge}}
\newcommand{\gh}{\text{\tiny gh}}
\newcommand{\gf}{\text{\tiny gf}}

\newcommand{\cl}{\text{\tiny cl}}
\newcommand{\diff}{\text{\tiny diff}}
\newcommand{\RG}{\text{\tiny RG}}

\begin{document}

\title{Effective universality in quantum gravity}

\author{Astrid~Eichhorn}
\affiliation{Institut f\"ur Theoretische Physik, Universit\"at Heidelberg,
Philosophenweg 16, 69120 Heidelberg, Germany}
\author{Peter~Labus}
\affiliation{International School for Advanced Studies, via Bonomea 265, 34136 Trieste, Italy}
\author{Jan~M.~Pawlowski}
\affiliation{Institut f\"ur Theoretische Physik, Universit\"at Heidelberg,
Philosophenweg 16, 69120 Heidelberg, Germany}
\affiliation{ExtreMe Matter Institute EMMI, GSI Helmholtzzentrum für
Schwerionenforschung mbH, Planckstr.\ 1, 64291 Darmstadt, Germany}
\author{Manuel~Reichert}
\affiliation{Institut f\"ur Theoretische Physik, Universit\"at Heidelberg,
Philosophenweg 16, 69120 Heidelberg, Germany}

\begin{abstract}
  We investigate the asymptotic safety scenario for a scalar-gravity
  system. This system contains two avatars of the dynamical Newton
  coupling, a gravitational self-coupling and a scalar-graviton
  coupling.  We uncover an effective universality for the dynamical
  Newton coupling on the quantum level: its momentum-dependent 
  avatars are in remarkable quantitative agreement in the scaling regime of 
  the UV fixed point. 
  For the background Newton coupling, this effective universality
  is not present, but qualitative agreement remains.
\end{abstract}

\maketitle

\section{Introduction}
In asymptotically safe quantum gravity the high-energy regime is
governed by a non-Gaussian fixed point. This fixed point renders the
ultraviolet (UV) behaviour finite, making the theory non-perturbatively 
renormalisable \cite{Weinberg:1980gg,Reuter:1996cp}. In recent years 
substantial evidence was collected in favour of this scenario
\cite{
  Christiansen:2012rx,Christiansen:2014raa,Christiansen:2015rva,Denz:2016qks,
  Meibohm:2015twa,Dona:2015tnf,Eichhorn:2017sok,Christiansen:2017cxa,
  Knorr:2017fus,Christiansen:2017bsy,Knorr:2017mhu,
  Manrique:2009uh,Manrique:2010mq,Manrique:2010am,Becker:2014qya,
  Reuter:2008wj,Reuter:2008qx,Donkin:2012ud,Dietz:2015owa,Safari:2015dva,
  Labus:2016lkh,Morris:2016spn,Percacci:2016arh,Ohta:2017dsq,Nieto:2017ddk,
  Lauscher:2001ya,Reuter:2001ag,Lauscher:2002sq,Litim:2003vp,Fischer:2006fz,
  Codello:2006in,Machado:2007ea,Codello:2008vh,Benedetti:2009rx,Eichhorn:2009ah,
  Eichhorn:2010tb,Groh:2010ta,Manrique:2011jc,Benedetti:2012dx,Dietz:2012ic,
  Falls:2013bv,Codello:2013fpa,Falls:2014tra,Becker:2014jua,Falls:2015qga,
  Eichhorn:2015bna,Falls:2015cta,Demmel:2015oqa,Gies:2015tca,Gies:2016con,
  Ohta:2016npm,Biemans:2016rvp,Ohta:2016jvw,Christiansen:2016sjn,Falls:2017cze,
  Gonzalez-Martin:2017gza,Wetterich:2017ixo,deAlwis:2017ysy,Falls:2017lst,
  Wetterich:2018qsl,Eichhorn:2018whv},
also in scalar-gravity systems 
\cite{Meibohm:2015twa,Dona:2015tnf,Eichhorn:2017sok,Dou:1997fg,Percacci:2002ie,
  Percacci:2003jz,Narain:2009fy,Eichhorn:2012va,Henz:2013oxa,Dona:2013qba,Percacci:2015wwa,
  Labus:2015ska,Oda:2015sma,Henz:2016aoh,Wetterich:2016uxm,Biemans:2017zca,Hamada:2017rvn,
  Becker:2017tcx,Eichhorn:2017als,Alkofer:2018fxj}.
See, e.g., 
\cite{Niedermaier:2006ns,Litim:2011cp,Reuter:2012id,Bonanno:2017pkg,
  Percacci:2017fkn,Eichhorn:2017egq}
for reviews.

The renormalisation group (RG)-idea behind asymptotic safety requires
the split of the full metric $g_{\mu\nu}$ into a background metric
$\bar g_{\mu\nu}$ and a dynamical fluctuation field $h_{\mu\nu}$ that
carries the quantum fluctuations. Naturally, one of the most important
ingredients in this approach is keeping track of diffeomorphism
invariance and background independence. In the present work we address
two questions that are linked to these key properties:

The first question concerns the dynamical couplings of gravity-matter
systems. Gauge theories feature different \emph{avatars} of the gauge
coupling, a prominent example being the running gauge couplings in the 
Standard Model. For instance, in QED the running electric coupling can 
be extracted from the wave function of the photon or from the 
electron-photon vertex. This relation follows from the Ward identities 
in QED. In QCD the running of the gauge coupling can be extracted from 
different combinations of vertex and propagator scalings including, 
e.g., the three-gluon vertex and the quark-gluon vertex. Again this can 
be derived from the identities following from the gauge symmetry, in 
this case the Slavnov-Taylor identities.

In these examples the respective couplings are marginal and exhibit
two-loop universality. This facilitates the identification. In gravity
the above universality holds for the (marginal) $R^2$ and
$R_{\mu\nu}^2$ couplings. However, the couplings in the classical
Einstein-Hilbert action and the minimal couplings to matter are
dimensionful and universality is not expected on the quantum
level anymore. Still, the multi-graviton couplings related to Taylor
expansions of the terms in the classical action, e.g.\
$1/G_N \sqrt{\det g}\, R$ and $1/G_N \sqrt{\det g}\, \Lambda$ agree on
the classical level and are related by Slavnov-Taylor identities in
quantum gravity. It is an intriguing physics question and of paramount
technical importance, whether for all practical purposes these
relations facilitate an identification, for example,
between all \emph{avatars} of the minimal coupling in gravity, the
Newton coupling $G_N$. We call this scenario \emph{effective
  universality}, which is detailed in the next section. 
The quest for effective universality is motivated by
its underlying physics properties. If realised it would hint at a
near-perturbative nature of the asymptotically safe fixed point
and, more importantly, at the physical nature of the fixed point as 
it is unlikely that a truncation artefact exhibits this property. 

In the present
work we investigate this question focusing on the dynamical pure
gravity coupling and the dynamical gravity-scalar coupling.  We stress
that due to the dimensionful nature of the couplings, different
\emph{avatars} of the Newton coupling could agree if evaluated within
the same scheme, but will of course depend, e.g., on the choice of
regulator function in the context of an functional RG setup. In this
work, universality is not to be understood in the sense of
scheme-independence at the two-loop level.

We find that the fixed-point 
values and the leading coefficients of the above two couplings
agree on a semi-quantitative level as a function of the number of
minimally coupled scalars.
These computations of dynamical couplings in a vertex expansion about
a flat background extend the previous works of
Refs.~\cite{Christiansen:2012rx,Christiansen:2014raa,
  Christiansen:2015rva,Denz:2016qks,Meibohm:2015twa,Christiansen:2017cxa,
  Dona:2015tnf,Eichhorn:2017sok}. For extensions about a curved background, see, 
e.g.~\cite{Christiansen:2017bsy, Knorr:2017fus}. 

The second question concerns the couplings of the background metric.
These couplings are related via Ward-identities. Additionally, they
are related to the dynamical couplings discussed above via Nielsen
identities or split Ward identities. Importantly, these identities
also carry the background independence of quantum gravity. In the
present work we investigate to what extent the
\emph{avatars} of the background couplings can be identified with that
of the fluctuation couplings. We emphasise that this identification is
at the root of the background-field approximation whose background
independence stands or falls with the validity of this
identification. 

In this work we do not only compare the avatars of the
background Newton coupling to that of the fluctuation coupling, we
also further improve the background coupling to a level-one coupling
with the explicit use of a Nielsen identity. Related works on Nielsen
identities in quantum gravity are
\cite{Reuter:2008qx,Reuter:2008wj,Donkin:2012ud,
  Dietz:2015owa,Safari:2015dva,Labus:2016lkh,Morris:2016spn,
  Percacci:2016arh,Ohta:2017dsq,Nieto:2017ddk}.  We find that the
effective universality that exists between the dynamical couplings
is not present for the background and the level-one couplings.  However,
general qualitative features of the flow equations, such as as the
sign of the scalar contribution, are preserved for all avatars of 
the Newton coupling.

\section{Avatars of couplings and effective universality}
\label{sec:avatun}
In this section we explain the origin of different avatars of couplings in
the effective action of matter-gravity systems. We further discuss their relation via
the modified symmetry relations, STIs and Nielsen-identities, that are
derived from the underlying diffeomorphism invariance and its breaking
in the presence of cutoff terms. In short, effective universality is
the notion that these complicated symmetry identities are well
approximated by Ward identities,
that is a diffeomorphism-invariant
approximation of the effective action, for more details see
\autoref{sec:effu}.

\subsection{Avatars of couplings in matter-gravity systems}
\label{sec:avatars}
Asymptotically safe matter-gravity systems and their physics can be
described in terms of the effective action
$\Gamma[\bar g_{\mu\nu},h_{\mu\nu},\varphi]$ in a gauge-fixed
setting. In the effective action we have dropped the Faddeev-Popov ghosts and
restricted ourselves to the case with scalar matter described by
$\varphi$. The field $h_{\mu\nu}$
denotes dynamical metric fluctuations around a generic background
metric $\bar g_{\mu\nu}$. The occurrence of the latter comes hand in
hand with the gauge fixing. The relation between background metric and
fluctuation field is not necessarily linear, but in the present work
we consider the linear split
\begin{align}\label{eq:linpara}
  g_{\mu \nu}=\bar g_{\mu \nu}+\sqrt{Z_{h} G_N}\, h_{\mu \nu} \,,
\end{align}
with the graviton wave-function renormalisation $Z_h$ normalised to one at some RG
scale $\Lambda$. In \eqref{eq:linpara} we have dropped the wave-function
renormalisations of $g_{\mu\nu}, \bar g_{\mu\nu}$ for the sake of
readability. The $G_N$-factor leads to a fluctuation field
$h_{\mu\nu}$ with the canonical dimension one. The effective action
$\Gamma[\bar g_{\mu\nu},h_{\mu\nu},\varphi]$ is not
diffeomorphism-invariant, but satisfies non-trivial Slavnov-Taylor
identities (STIs). The information on physics is carried by the
diffeomorphism-invariant effective action, 
\begin{align}\label{eq:BackAc}
  \Gamma[g_{\mu\nu},\varphi]= \Gamma[g_{\mu\nu},h_{\mu\nu}=0,\varphi]\,.
\end{align}
and the respective background correlation functions are related to
S-matrix elements. However, their computations requires the knowledge
of the scattering processes with dynamical gravitons
$h_{\mu\nu}$. These processes are described through different couplings
$G_{\vec n}$, where the vector $\vec
n$ consists of the numbers of the different dynamical fields $\Phi =
\big( h_{\mu\nu},\,c_{\mu},\,\varphi,\,\dots
\big)$ that take part in the process,
\begin{align}
  \vec n = \big( n_h,\, n_c,\,n_\varphi,\,\dots \big) \,.
\end{align}
The couplings $G_{\vec n}$ of all these fields to the dynamical
graviton are avatars of the gravitational self-coupling $G_N$. In the
present study we concentrate on gravity-scalar couplings. This leaves
us with couplings labelled by two indices only,
\begin{align}\label{eq:coupling}
G_{(n_h,n_\varphi)} \,.
\end{align}
We denote dimensionless versions of the
Newton coupling by capital letters, e.g.~as above
$G_{(n_h,n_\varphi)}$, and dimensionful versions with an additional
over-bar, such as~$\bar G_{(n_h,n_\varphi)}$. The fluctuation couplings
$G_{(n_h,n_\varphi)}$ defined in \eqref{eq:coupling} are related to the
expansion coefficients in an expansion in powers of $h_{\mu\nu}$, the couplings
$G_{(n_{\bar g},n_\varphi)}$ are related to those in an expansion in
powers of $\bar g_{\mu\nu}$.  The respective vertices are given by
\begin{align}\nonumber 
 &\Gamma^{(n,m,l)}(p_1,\ldots,p_{n+m+l})\\[1ex]
 &\qquad=\0{\delta^{n+m+l}\Gamma[\bar g,h,\varphi]}
    {\delta\bar g^n(p_1,\ldots)\delta h^m(\ldots,p_i,\ldots)\delta\varphi^l(\ldots,p_{n+m+l})}\,, 
  \label{eq:Gnm}
\end{align}
where we suppress the indices on $\bar g$ and $h$ for brevity of
notation. The couplings \eqref{eq:coupling} are now defined by
\eqref{eq:Gnm} at selected kinematic configurations. In this work, we
focus on $G_{(3,0)}$ and $G_{(1,2)}$ defined at the momentum-symmetric
point. The former coupling relates to the scattering of three
gravitons, and is derived from the pure gravity part of the effective
action. The coupling $G_{(1,2)}$ relates to the scattering of one
graviton and two scalars, and is derived from the kinetic term of the
scalars. This coupling is also present in the free -no
self-interaction- scalar theory and can be considered the fundamental
coupling of scalar fields to gravity. 
We compare these couplings based on their flow equations. We do not 
compute the explicit STIs that relate them. In general the information
from the flow and the STIs is equivalent, but in a truncated 
non-perturbative computation they will not agree. It is an important task 
to quantify to what extent the STIs are satisfied but this goes 
beyond the scope of this work. The computation of the STIs is a technically challenging
task, for an example in QCD see \cite{Ellwanger:1995qf,Cyrol:2017ewj}.

Evidently these couplings cannot be defined uniquely and depend on the
given kinematical limit. Note that this even holds for dimensionless
couplings beyond one loop, despite their universal RG
running. Accordingly, the evaluation of, e.g., scattering processes
with different momentum configurations requires an analysis of the
corresponding $n$-point vertex as a function of all its independent
momenta, i.e., a simple function of one momentum cannot capture the
full dynamics adequately.  If dealing with an approximation to the
theory that does not maintain the full momentum-dependence of
vertices, a typical choice is the symmetric point, for higher-order
vertices \emph{a} symmetric point. Using these momentum configurations
can lead to semi-quantitative agreement with the full results even in
strongly-correlated systems, for recent work in gravity see
\cite{Denz:2016qks}, and in QCD see \cite{Cyrol:2017ewj}. For a
related discussion in the effective field theory approach
to gravity see \cite{Anber:2011ut}.  Keeping this caveat in mind, we
proceed with our study whether avatars of the Newton coupling,
defined using the symmetric momentum configuration of various
three-point vertices, show semi-quantitative agreement.

With the dynamical vertices \eqref{eq:Gnm} and the dynamical propagators
we can compute the background vertices, that is the $S$-matrix
elements. This leads to further avatars of the Newton coupling, this
time being directly related to S-matrix elements for the selected
momentum configuration. In the present work we consider the avatar of the
Newton coupling of the background curvature term in the
action. It is distinguished from the $G_{\vec n}$ by two properties:
first it is the prefactor of a diffeomorphism invariant term in the
action. Second, as a pure background quantity it does not drive the RG
flow of the system, which is driven by the fluctuation field and its
couplings. In this work, we refer to its dimensionless version as
$\bar G$ and the dimensionful version as $G_N$.

\subsection{RG-approach to asymptotically safe matter-gravity systems}
\label{sec:FRGformal}
The standard approach to computations in asymptotically safe gravity
is the functional renormalisation group (FRG). The FRG approach to quantum
gravity is based on the flow equation for the effective action, the Wetterich equation 
\cite{Wetterich:1992yh, Ellwanger:1993mw,Morris:1993qb},
\begin{subequations} 
\label{eq:flow-master} 
\begin{align}
  \partial_t \Gamma_k = 
\frac{1}{2}{\rm Tr} \left[G_k\, \partial_t R_k \right]\,. 
\label{eq:flow}
\end{align}
In \eqref{eq:flow} we have introduced the RG-time $t = \log k/k_0$ with a reference scale 
$k_0$. The trace sums/integrates over the
discrete/continuous spectrum of the propagator $G_k$.  We emphasise that the
flow of the effective action is solely driven by the second
derivatives of the effective action w.r.t.\ the fluctuation
fields. Accordingly it depends on the fluctuation propagators
\begin{align}
  \label{eq:prop} 
  G_k= \0{1}{\Gamma_k^{(0,2)}+R_{k}}\,, 
 \end{align}
\end{subequations}
with the second fluctuation field derivatives defined in
\eqref{eq:Gnm}, and the background-metric-dependent graviton and
scalar field regulators $R_{h,k}$ and $R_{\varphi,k}$ respectively.
While $\Gamma_k^{(0,2)}$ is a matrix in field space, which features
off-diagonal components, the regulators are chosen to be diagonal. We
have dropped the ghost contribution in \eqref{eq:flow} for the sake of
readability while taking it into account in our calculations.
Induced scalar-ghost interactions \cite{Eichhorn:2013ug} are neglected in our
truncation.

Let us now come back to the question of the symmetry identities
mentioned in \autoref{sec:avatars}. While background
diffeomorphism invariance is introduced as a mere computational tool
and can even be established for theories without diffeomorphism
invariance, it inherits the physical diffeomorphism invariance of
gravity via the diffeomorphism STIs and the NIs/sWIs. The latter carry
the background independence of gravity by relating derivatives with
respect to the background metric $\bar g_{\mu\nu}$ and the fluctuation
field $h_{\mu\nu}$.

In summary, the approach encodes the background independence and
diffeomorphism invariance of observables in a counter-intuitive way:
background independence of the setting implies nontrivial relations
instead of simple equalities between couplings that would be equal in
a classical, diffeomorphism invariant setting. This also implies that
diffeomorphism invariant approximations to
$\Gamma[\bar g_{\mu\nu}, \Phi]$ are potentially at odds with
physical diffeomorphism invariance and background independence. They
should be taken with a grain of salt and have to be investigated
thoroughly. The current work is a first step in this direction in a
coupled matter-gravity system.

In the present renormalisation group setup the situation is even more
intricate as diffeomorphism invariance and background independence are
broken by the presence of the infrared regularisation. Any local
coarse graining procedure requires the introduction of a background in
order to define a notion of high-momentum modes.  The presence of the
corresponding cutoff term leads to \emph{modified} STIs, NIs/sWI.  In
the limit $k\to\infty$ these deviations from the standard STI and sWI
may play a crucial r\^ole for the correct description of the physical
dynamics. In order to restore background independence in the physical
limit $k \rightarrow 0$, the violation of diffeomorphism invariance
and background independence introduced via the regulator must be
compensated for by an appropriate UV initial condition. This UV
initial condition violates diffeomorphism invariance and background
independence such that the violation is fully 'eaten up' by the RG
flow to the IR. Note also that the \emph{physical} UV limit is the one
where physical scales (momenta, curvature etc) take large values, but
$k$ is kept at $k=0$. Although physical scales can act as an IR
cutoff, the UV limit $k \rightarrow \infty$ could show differences
from the limit where physical scales such as momentum or curvature 
scales take their UV limit.

\subsection{Effective universality}
\label{sec:effu}
In the setup in \autoref{sec:avatars} and \autoref{sec:FRGformal}
already one diffeomorphism-invariant operator at the classical level,
for example the curvature scalar $\sqrt{g}R$ leads to
infinitely many different couplings at the quantum level: These are
obtained by taking the $n$th $h_{\mu\nu}$-derivative of
$\sqrt{g}R$ and projecting $\Gamma^{(n)}$ (given a complete
basis) on this tensor structure. While still being related by STIs
their couplings do not agree. In the presence of the regularisation these STIs
turn into mSTIs.

The situation is slightly different for the $n$th order
background couplings: they even agree at the full quantum 
level as they are related by
Ward identities due to background diffeomorphism
invariance. This property even survives the introduction of the
regularisation. However, as discussed above, 
the computation of their $\beta$-functions requires
the knowledge of the fluctuation vertices. They 
are related to the background vertices by the
Nielsen or split Ward identities, which turns the Ward identities into
the STIs. In the presence of the regularisation we have modified NIs 
as we have mSTIs.  

This leaves us with the technical challenge of computing all these
coupling avatars related to a given operator, in the present example
the avatars of the Newton coupling. Specifically, the challenge lies
in the need to close a given system of flow equations for correlation
functions that depend on the higher-order correlation functions. To
that end one has to provide an ansatz for higher-order couplings for
which the flow is not computed.  The canonical choice is their
classical value. In quantum gravity this canonical choice leads to an
identification of all higher-order couplings derived from a given
operator with the lowest-order one, effectively restoring
diffeomorphism invariance. This we call \emph{effective universality}.

For example, let us assume for a moment that we only compute the flow
of one avatar of the Newton coupling.  Then the canonical choice leads
to the identification of all higher order Newton couplings with the
lowest order one. If we apply this concept to the dynamical system,
effective universality can be summarised by
\begin{align}\label{eq:fluc-effu} 
G_{(n_h,n_\varphi)} \approx G\,,
\qquad 
n_h,n_\varphi \in\mathbb{N}\,,
\end{align}
with a unique Newton coupling for a suitably chosen momentum configuration.
One of the main aims of this paper is to compare the scale dependence
of these couplings under the impact of quantum fluctuations of the
metric and of $N_s$ scalar fields. 

In its maximal version for both, background couplings and fluctuation 
couplings, effective universality can be summarised in a
concise form of the effective action,
\begin{align}\nonumber 
  \Gamma[\bar g_{\mu\nu},h_{\mu\nu},\varphi] &= 
  \Gamma_\diff [\bar g_{\mu\nu}+h_{\mu\nu},\varphi]\\[1ex] 
  &\qqquad+\Delta \Gamma_\gauge [\bar g_{\mu\nu},h_{\mu\nu},\varphi]\,.
\label{eq:backsplit}
\end{align}
with a diffeomorphism-invariant action $\Gamma_\diff[g]$ and
\begin{align}\label{eq:gaugesplit}
  \Delta \Gamma_\gauge [\bar g_{\mu\nu},h_{\mu\nu},\varphi]
  \approx S_\gf [\bar g_{\mu\nu},h_{\mu\nu}]
  +S_\gh[\bar g_{\mu\nu},h_{\mu\nu},c_{\mu}]\,,
\end{align}
with gauge fixing and ghost action, $S_\gf$ and $S_\gh$, respectively,
see \cite{Reuter:1993kw,Reuter:1996cp}. Furthermore, only the
regulator terms would carry the breaking of background independence.
This approximation is called the \emph{background-field approximation}.
It has been used predominantly in the RG approach to quantum gravity 
and is being paramount to effective field theory applications in 
quantum gravity \cite{Donoghue:1993eb,Robinson:2005fj,Pietrykowski:2006xy,
Toms:2007sk,Ebert:2007gf,Rodigast:2009zj,Alvarez:2015sba}. 

In summary the quest for \emph{effective universality} is
directly related to the task of finding an efficient (rapidly
convergent) expansion of the quantum effective action of
matter-gravity systems in diffeomorphism-invariant operators. While
this task is seemingly a technical one it is -in disguise- the quest 
for the aspects of physics that govern quantum gravity systems.

\section{RG for scalar-gravity systems}
In the present work we aim to shed light on the above issues and
specifically explore in which settings effective universality may emerge
in simple approximations. To that end we compare two avatars of the dynamical 
Newton coupling in this section.
The first is defined from the three-graviton vertex, as in
\cite{Christiansen:2015rva,Meibohm:2015twa,Denz:2016qks,Christiansen:2017cxa}, 
and its dimensionless version is called $G_{(3,0)}$. The second one is defined
from the graviton-two-scalar vertex as in \cite{Dona:2015tnf,Eichhorn:2017sok}, 
and is called $G_{(1,2)}$.

To project the RG flow \eqref{eq:flow} onto a given coupling
$G_{(n,m)}$ we take $n$ functional derivatives with respect to the
graviton and $m$ with respect to the scalar field.  The resulting
tensor has $2n$ open indices that we contract with an appropriate
tensor structure, see
\cite{Christiansen:2015rva,Meibohm:2015twa,Denz:2016qks,Christiansen:2017cxa}
and \cite{Dona:2015tnf,Eichhorn:2017sok} for details.  The resulting
structure then depends on the momenta of the $n+m$ external legs of
the respective couplings.  For the two couplings mentioned above, we
use a symmetric momentum configuration, where we set the angles
between each pair of momenta to $2\pi/3$ and the magnitude of the
momenta to $p$.  The couplings in a vertex expansion are actually
momentum-dependent functions.  For the analytic results, we project at
$p=0$ to instead model them by a single number.  For the numerical results,
we instead utilise a bilocal projection at $p=0$ and $p=k$ as in 
\cite{Christiansen:2015rva,Meibohm:2015twa,Denz:2016qks,Christiansen:2017cxa}:
This projection is motivated by the fact that the momentum integrals
are peaked at $p\approx k$. Hence this momentum regime is more
important for quantitative accuracy and partially also for capturing
qualitative features of the flow of correlation functions.  The
bilocal projection partially captures the global momentum
dependence. It has been tested successfully against the fully
momentum-dependent results in
\cite{Christiansen:2015rva,Meibohm:2015twa,Denz:2016qks,Christiansen:2017cxa}. 

To define our truncation, we start from an Einstein-Hilbert
action that is accompanied by a gauge fixing and ghost action,
and a canonical kinetic term for the scalar fields,
\begin{align}
  S ={}& -\01{16 \pi G_N} \int \mathrm d^4 x\, \sqrt{g}\left(R
        - 2 \Lambda \right)+ S_\gf + S_\gh \notag\\[1ex]
    &+\012 \, \sum_{i=1}^{N_s} \int \mathrm  d^4 x\, \sqrt{g}
         g^{\mu \nu} \partial_{\mu}\varphi^i\partial_{\nu}\varphi^i \,.
\end{align}
We use the standard Faddeev-Popov gauge fixing procedure, with gauge fixing action
\begin{align}
  S_\gf &= \01{2\alpha} \,
    \int \mathrm d^4x \sqrt{ \bar{g} } \, F_\mu \bar g^{\mu\nu} F_\nu \,, \notag\\[1ex]
  F_\mu &= \bar \nabla ^\nu h_{\mu\nu} - \0{ 1 + \beta }4 \,\bar \nabla_\mu h^{\nu}{}_{\nu}\,.
\end{align}
We specialise the background metric to a flat Euclidean one,
$\bar g_{\mu\nu}  = \delta_{\mu\nu}$,
and work with the values $\alpha=0$ and $\beta=1$ for the gauge
parameters. This is a fixed point of the RG flow \cite{Litim:1998qi}, as are all 
combinations $\alpha=0$ and $\beta$.
This choice of gauge parameters is technically favourable on a flat 
background since the poles of all modes of the classical graviton 
propagator coincide. Later in this work, for the level-one 
improvement, we also resort to the gauge choice $\alpha=\beta=0$.

Next we insert the linear parameterisation \eqref{eq:linpara} into
the above action and subsequently expand the Einstein-Hilbert action
up to fifth order and the kinetic term of the scalar up to the third order 
in the fluctuation field $h_{\mu\nu}$. Taking into account that 
all field monomials in the action evolve independently under the RG
dynamics due to the breaking of diffeomorphism invariance,
we introduce a separate dimensionful coupling for each vertex and denote
it by $\bar{G}_{(n_h, n_{\varphi})}$. We also distinguish between
different avatars of the cosmological constant, introducing
a dimensionful graviton mass parameter $\bar{\mu}$
associated to a mass-like term in the graviton propagator,
and couplings $\bar{\lambda}_n$ associated to 
the momentum-independent part of the $n$-graviton vertex.
In our approximation, the
vertex functions are written as Einstein-Hilbert tensor structures
with the appropriate substitutes of the cosmological and Newton
constant
\begin{align}\label{eq:vert}
  \Gamma_k^{(0,n,m)}=
  S^{(0,n,m)}(\mathbf{p};
  \Lambda \to \Lambda_n,
  G_N \to \bar{G}_{(n,m)})\,.
\end{align}
Here, $\mathbf{p}=(p_1,\ldots,p_{n+m})$ denotes the momenta of the external fields.
Note that the pure gravity terms in \eqref{eq:vert} 
are proportional to $\bar G_{(n,0)}^{n/2-1}$ while the gravity-matter terms are
proportional to $\bar G_{(n,m>0)}^{n/2}$.
Furthermore, \eqref{eq:vert} is proportional to $Z_{h}^{n/2} Z_{\varphi}^{m/2}$
due to the rescaling of the fluctuation fields, see \eqref{eq:linpara}. 
This construction together with the choice of regulator assures that the
wave-function renormalisations only enter via
the corresponding anomalous dimensions $\eta_{i}$. These are defined via
\begin{align} \label{eq:def-anom-dim}
	\eta_{i}(p^2) := -\partial_t \ln Z_{i}(p^2)\,.
\end{align}

Schematically the scale dependent action reads
\begin{align}\label{eq:vexp}
  \Gamma_k[\bar g, h,\varphi]
  ={}& \, \Gamma^{(0,0,0)}_k[\bar{g}] + \Gamma^{(0,1,0)}_{k}[\bar{g}] h
   \notag \\[1ex]
  & + \012 \Gamma^{(0,2,0)}_{k}[\bar{g}] h^2 
  + \01{3!}\Gamma^{(0,3,0)}_{k}[\bar{g}] h^3  \\[1ex]
  & + \012 \Gamma^{(0,0,2)}_{k}[\bar{g}] \varphi^2
  + \012 \Gamma^{(0,1,2)}_{k}[\bar{g}] h \varphi^2
  + \ldots \,, \notag
\end{align}
where we suppress contributions coming from ghost fields to improve
readability. The ghost two-point function and the ghost contributions
to the running of other $n$-point functions are taken into
account in our work.

In truncations one challenge is to find a way to
consistently close the infinite tower of flow equations for the vertices. Such
a consistent closure is expected to lead to an enhanced robustness of
the truncation. Specifically the flow of an $n$-point vertex depends on the $n+2$-point vertices.
We explicitly evaluate the flow of couplings with $n \leq 3$ and
equate the higher-order couplings $G_{(4,0)}$, $G_{(5,0)}$, 
$G_{(2,2)}$ and $G_{(3,2)}$ as well as $\lambda_4$ and 
$\lambda_5$ with the lower-order couplings.

In summary we compute and evaluate the coupled flow equations of the 
scale dependent dimensionless quantities
\begin{align}
 \bar \lambda,\  \bar G,\  \mu,\  \lambda_3,\  G_{(3,0)},\, 
 G_{(1,2)},\  \eta_h(p^2),\ \eta_\varphi(p^2) ,\ \eta_c(p^2) \,.
 \label{eq:full-system}
\end{align}
The background couplings $\bar \lambda$ and $\bar G$ do not enter the
flow and thus do not affect the fluctuation couplings. Their flow
equations are analytic and derived using the York-decomposition
\cite{York:1973ia,Stelle:1976gc} with field redefinitions
\cite{Dou:1997fg,Lauscher:2001ya}.  The explicit pure gravity flow
equation for our gauge is displayed in
\cite{Gies:2015tca,Denz:2016qks}.  The $N_s$-dependent part is gauge
independent and thus equal to, e.g.~\cite{Dona:2013qba}.  The flow
equations for $\mu$ and $\lambda_3$ are also analytic and agree with
\cite{Meibohm:2015twa} (the coupling $G_{(1,2)}$ has to be
disentangled from $G_{(3,0)}$ in the appropriate terms).  The
momentum dependence of the Newton couplings, $G_{(3,0)}$ and
$G_{(1,2)}$, and the anomalous dimensions, $\eta_h$ and
$\eta_\varphi$, is of importance and thus, following the discussion in
\cite{Christiansen:2015rva,Meibohm:2015twa,Denz:2016qks,Christiansen:2017cxa},
it is preferable to evaluate these at finite momentum, which does not
allow for analytic equations. Nevertheless the analytic version of
these flows leads to qualitatively reliable results. The analytic and momentum-dependent
versions of $G_{(3,0)}$, $\eta_h$, and $\eta_\varphi$ agree
with \cite{Meibohm:2015twa}. Again $G_{(1,2)}$ has to be
distinguished from $G_{(3,0)}$. The analytic version of
$G_{(1,2)}$ is the same as in \cite{Eichhorn:2017sok} while the
momentum-dependent version is derived for the first time in this work. 

The bilocal projection prescriptions for the Newton couplings lead to
\begin{align}
  \beta_{G_{(3,0)}} ={}& \left(2 + 3 \eta_h(k^2) \right)G_{(3,0)} \notag\\[1ex]
                       &-\0{24}{19}\left(\eta_h(k^2)
                         -\eta_h(0)\right)\lambda_3\, G_{(3,0)} \notag\\[1ex]
                       &+ (32\pi)^2 \0{64}{171}\,G_{(3,0)}^{1/2} \notag\\[1ex]
                       &\quad \times\Bigl(\Flow^{(hhh)}_{{\TT},G_{(3,0)}}(k^2)-
                         \Flow^{(hhh)}_{{\TT},G_{(3,0)}}(0)\Bigr)\,,\notag\\[1ex]
  \beta_{G_{(1,2)}} ={}& \left(2 + \eta_h(k^2) + 2 \eta_\varphi(k^2)\right) G_{(1,2)} \notag\\[1ex]
                       &+ \0{8}{3}\, G_{(1,2)}^{1/2}\,
                         \Flow^{(h\varphi\varphi)}_{{\TT},G_{(1,2)}}(k^2)\,.    
  \label{eq:flow-eqs-g}
\end{align}
Here, the notation $\TT, G_{(3,0)}/G_{(1,2)}$ indicates a contraction
of the right-hand side of the Wetterich equation with the projection
operator onto the corresponding coupling as in
\cite{Meibohm:2015twa,Denz:2016qks}. By
$\Flow^{(hhh)}_{{\TT},G_{(3,0)}}(k^2)$ we refer to the right-hand side
of the Wetterich equation, projected on three external $h_{\mu\nu}$
legs, each contracted with a transverse projector, and evaluated at
the external momentum set to $p^2=k^2$. The prefactors such as
$\0{24}{19}$ or $(32\pi)^2 \0{64}{171}$ arise from the contraction of the
respective tensor structures with the projection operators.  For
$\beta_{G_{(3,0)}}$ they are identical to
\cite{Christiansen:2015rva,Meibohm:2015twa,Denz:2016qks} and for
$\beta_{G_{(1,2)}}$ to \cite{Dona:2015tnf} up to a rescaling of
$h_{\mu\nu}$.  The argument of $\Flow (x)$ denotes the magnitude of
the external momenta on the right-hand side, which are set to the
momentum-symmetric point. Thus $ \Flow^{(hhh)}_{{\TT},G_{(3,0)}}(0)$
is the right-hand side of the Wetterich equation projected onto
contributions with three external gravitons and evaluated at vanishing
external momentum. Due to the shift symmetry in the scalar sector,
$\Flow^{(h\varphi\varphi)}_{{\TT},G_{(1,2)}}(0)=0$.  Note further that
we compute the momentum-dependent anomalous dimensions with the
approximation that we evaluate the anomalous dimension at $p^2=k^2$ if
it appears in an integral, see \cite{Meibohm:2015twa} for a discussion
of this approximation.

For the derivation of e.g.~the Flow-expressions in
\eqref{eq:flow-eqs-g} we used the symbolic manipulation system 
{\small \emph{FORM}}~\cite{Vermaseren:2000nd,Kuipers:2012rf} as well
as the \emph{FormTracer}~\cite{Cyrol:2016zqb} to trace diagrams.

\section{Effective universality for the dynamical couplings}
\label{sec:uni}
To address our first key question, we compare the $\beta$-functions and
fixed-point results for the dynamical system including
$G_{(3,0)},\, G_{(1,2)},\, \mu$ and $\lambda_3$. 
The $\lambda_n$ and in particular $\mu= -2 \lambda_2$ play a
special r\^ole due to the convexity of the effective action. To see
this consider the effective action for classical gravity. It is the
double Legendre transform of the classical action. Accordingly, for
positive cosmological constant it only agrees with the classical
action for large enough curvature. Thus, even for a
diffeomorphism-invariant action, the $\lambda_n$ are not necessarily
the same. In summary, in the reduced system under investigation
effective universality may only hold directly for $G_{(3,0)}$ and
$G_{(1,2)}$ even if it is fully present. This leaves us with the
two avatars of the Newton coupling while $\mu$ and $\lambda_3$ should
be evaluated in dependence of $G_{(3,0)}$ and $G_{(1,2)}$ on a given
trajectory.

Effective universality is necessarily broken at a
finite cutoff scale as the regulators break diffeomorphism
invariance. Accordingly it cannot hold quantitatively for all cutoff
scales. It may hold at $k\to 0$, and potentially at $k\to \infty$.
While the former physical case is evident, the latter case deserves
some explanation: in the physics limit at $k=0$ and for momentum
scales $p\gg M_{\text{\tiny{Planck}}}$ we are in the scaling regime
of the UV fixed point. If effective universality holds for $k=0$,
we have in particular $G_{(3,0)}(p^2)\approx G_{(1,2)}(p^2)$. If we now
increase the cutoff scale, the scaling couplings are only changed for
$p^2\approx k^2$, i.e.\ the flows are local in momenta.
Accordingly, the high-momentum behaviour of $G_{(i,j)}(p^2)$ can only
be changed at large $k$, where one already probes the scaling regime. 
Then, self-similarity in the scaling regime
entails that $k$-scaling and $p$-scaling agree. Hence, \emph{physical}
effective universality at $k=0$ translates into effective universality
of the cutoff-dependent couplings in the scaling regime.

We shall see that $G_{(3,0)}$ and $G_{(1,2)}$ indeed feature a
semi-quantitative effective universality on scaling trajectories close to
the UV fixed point.

\subsection{Effective universality at the fixed point}
\label{sec:uni-fp}
We solve the flow equations of the fully coupled fluctuation system,
$G_{(3,0)},\, G_{(1,2)},\, \mu$ and $\lambda_3$, identifying all
higher-order gravity couplings with $G_{(n\geq3,0)}= G_{(3,0)}$ and
$\lambda_{n\geq3}= \lambda_3$, and all graviton-scalar couplings with
$G_{(n,m\geq2)}=G_{(1,2)}$.  The $\beta$-functions of the Newton
couplings are given schematically in \eqref{eq:flow-eqs-g}.

\begin{figure}[t!]
\includegraphics[width=\linewidth]{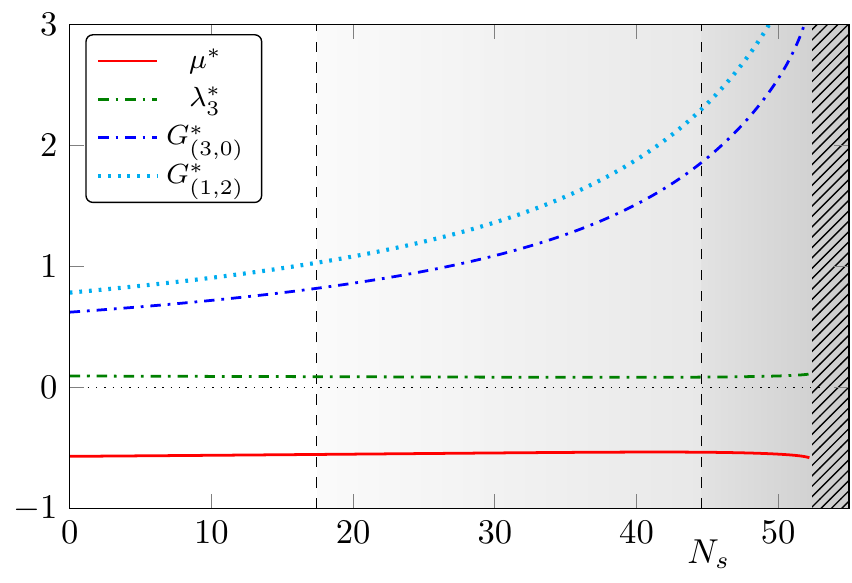}
\caption{Fixed-point values for the fluctuation couplings as a
  function of $N_s$.  The vertical lines at $N_s \approx 17.5$ and $N_s \approx 44.6$ show where
  $\eta_h(0)$ and $\eta_h(k^2)$ exceed two, respectively.}
\label{fig:FPvalues}
\end{figure}

The resulting fixed-point values are shown in
\autoref{fig:FPvalues}.  We observe that both Newton couplings have
similar fixed-point values that increase with $N_s$. The couplings
$\mu$ and $\lambda_3$ remain approximately constant as a function of
$N_s$. This already shows a qualitative effective universality for
$G_{(3,0)}$ and $G_{(1,2)}$ that supports the reliability of
computations where this property is used. The similar behaviour of the
two avatars of the Newton coupling for all $N_s$ and the
$N_s$-independence of $\mu$ and $\lambda_3$ suggests to first perform
a detailed analysis at a fixed $N_s$ and then a subsequent one of the
$N_s$-dependence. For the first part of the analysis we choose
$N_s=0$. This is in complete analogy to the quenched approximation in
QCD, where one drops all closed quark loops. In the present case it
amounts to dropping all closed scalar loops. This does not imply the
complete absence of scalar fluctuations, as they still appear in diagrams
with internal scalar and graviton lines, i.e.\ in the diagrams of the 
$G_{(1,2)}$ flow. 

\begin{figure*}[t]
\includegraphics[width=\linewidth]{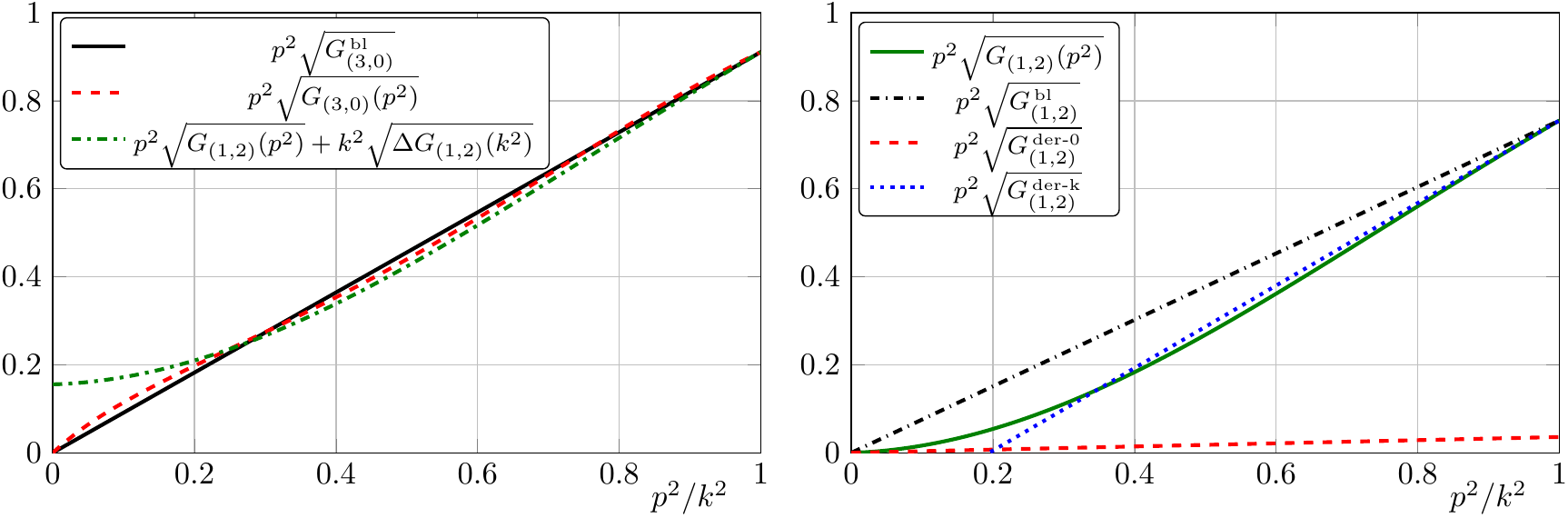}
\caption{Displayed are the momentum-dependent Newton couplings 
  evaluated on the quenched bilocal fixed point using effective universality for the
  momentum-independent couplings, see \eqref{eq:uniG}.
  Left: $p^2\sqrt{G_{(3,0)}(p^2)}$ and $p^2\sqrt{G_{(1,2)}(p^2)} - k^2\sqrt{\Delta G_{(1,2)}(k^2)}$, see \eqref{eq:offset12}. 
  The black lines guides the eye and corresponds to the bilocal approximation $p^2\sqrt{G^{\,\text{\tiny bl}}_{(3,0)}}$.
  Right: $p^2\sqrt{G_{(1,2)}(p^2)}$ as well as local and bilocal approximations to the momentum
  dependence of it.}
\label{fig:MomDep}
\end{figure*}

\subsection{Effective universality for quenched quantum gravity}
\label{sec:quenched}
In the quenched limit, $N_s=0$, a quantitative self-consistency
analysis reveals an even more interesting property than the mere
similarity observed in \autoref{fig:FPvalues}.  To that end we
remind ourselves that the bilocal projection used in the present 
fixed-point computation is based on observations in the pure gravity system
in \cite{Christiansen:2015rva,Denz:2016qks} for $G_{(3,0)}$ and
$G_{(4,0)}$. In particular in \cite{Denz:2016qks} it was shown that
the momentum dependence of the coupling $G_{(3,0)}$ and
$G_{(4,0)}$ related to the curvature term $R$ is quantitatively given
by a linear $p^2$-dependence. The four-graviton vertex has an
additional $p^4$-dependence related to the $R^2$-term, no 
higher-order momentum-dependence is present, for more details see
\cite{Denz:2016qks}. These properties are based on non-trivial
cancellations between diagrams based on diffeomorphism invariance.
This situation suggests the following self-consistency analysis of effective universality for
$G_{(3,0)}$ and $G_{(1,2)}$: assume that effective
universality works quantitatively at the fixed point, that is
\begin{align}\label{eq:uniG}
G_{\vec n}=G\,\qquad \text{with} \qquad G=G_{(3,0)}\,,
\end{align}
for all avatars of the Newton coupling. The self-consistency of
\eqref{eq:uniG} is tested quantitatively by evaluating the
momentum-dependent $\beta$-functions $\beta_{G_{(3,0)}}$ and
$\beta_{G_{(1,2)}}$ on \eqref{eq:uniG} and the approximately
$N_s$-independent fixed-point values $\mu^*$ and
$\lambda_3^*$. Solving the momentum-dependent $\beta$-functions, 
cf.~\eqref{eq:flow-eqs-g}, for the momentum-dependent couplings on 
the bilocal fixed-point values leads us to
\begin{align}
p^2\, \sqrt{G_{(3,0)}(p^2)} &\simeq 
- \0{64}{171}(32\pi)^2
\notag\\[1ex]
&\times\left.\0{\Flow^{(3,0)}(p^2)-\Flow^{(3,0)}(0) }{2+
3 \eta_h(p^2) }\right|_{G^*,\mu^*,\lambda_3^*}, \notag \\[1ex]
p^2\, \sqrt{G_{(1,2)}(p^2)} &\simeq 
- \083 \0{\Flow^{(1,2)}(p^2)}{2 + \eta_h(p^2) + 
2\eta_\varphi (p^2) }\bigg|_{G^*,\mu^*,\lambda_3^*}.
\label{eq:betap}
\end{align}
Note that $\eta_\varphi (p^2) = 0$ in the chosen gauge, $\beta=1$ and
$\alpha=0$.  The momentum-dependent fixed-point couplings are shown in
\autoref{fig:MomDep}.  There we have shifted $p^2\sqrt{G_{(1,2)}(p^2)}$ by a
constant in order to make the quantitatively coinciding
linear dependence for $p^2 \gtrsim 0.3\, k^2$ apparent, 
see \eqref{eq:inter12-crude} evaluated at $p^2=k^2$ for the definition of the shift.
This coincidence is a non-trivial consequence of the different
contributions of $\mu$ and $\lambda_3$ to both $\beta$-functions. It
entails effective universality on the quantitative level. The
deviation from effective universality at small momenta may have two
different sources: first we expect that the regulator-induced breaking
of effective universality is maximal at low momenta in comparison to the cutoff scale.
A second source of the deviation may be the graviton mass scale in the graviton
propagators, and could be related to the convexity-enforcement at work
in the effective action.

Typically the momentum dependence of $p^2\sqrt{G_{(1,2)}(p^2)}$ is
studied in terms of a derivative expansion about $p^2=0$.  For our
results this corresponds to a small $p^2$-term and large $p^4$ and
$p^6$-terms. This suggests an interpretation of the present results as
the dominant generation of higher-order couplings such as
$\sqrt{g}R_{\mu\nu}\varphi\nabla_\mu\nabla_\nu\varphi$.  On the other
hand, an expansion about a momentum $p^2 \gtrsim 0.3\, k^2$ leads to a
$p^2$-coefficient of the same order as in the three-graviton vertex,
which supports the emergence of effective universality. At the same
time the higher-order momentum coefficients of the expansion
$p^2 \gtrsim 0.3\, k^2$ are typically smaller by a factor three in
comparison to the leading one. The expansion about $p=0$ and its
interpretation as higher-order operators hinges on a
diffeomorphism-invariant expansion of the effective action deep in the
regime where the regulator spoils diffeomorphism-invariance.
Interestingly, the momentum dependence of $p^2\sqrt{G_{(3,0)}(p^2)}$
has a much clearer interpretation: the $p^2$-term is rather
independent of the projection scheme.

This discussion and the highly non-trivial result of quantitative 
effective universality in \autoref{fig:MomDep} suggests to take 
a closer look at how well local and bilocal approximations capture 
the full momentum dependence and how they accommodate effective 
universality. Accordingly, we close this section on quenched quantum 
gravity with a discussion and evaluation of different approximation 
schemes.

\subsubsection{Quantitative bilocal schemes}
\label{sec:effuni-quant}
The full momentum dependences of the Newton couplings are well
approximated by using the bilocal result for the three-graviton
coupling, $G^{\,\text{\tiny{bl}}}_{(3,0)}$, and by
$G^{\,\text{\tiny{bl}}}_{(3,0)}$ with an additional interpolating
piece for the scalar-graviton coupling.  This amounts to
\begin{subequations} 
\label{eq:offset12}
\begin{align}\label{eq:offset12G} 
p^2\sqrt{G^{\,\text{\tiny{quant}}}_{(1,2)}(p^2)}:=
p^2\sqrt{G^{\,\text{\tiny{bl}}}_{(3,0)}} + k^2\sqrt{\Delta  G_{(1,2)}(p^2)}\,, 
\end{align} 
with 
\begin{align} 
\sqrt{\Delta  G_{(1,2)}(p^2\gtrsim 0.3\, k^2)} &=
 \sqrt{G_{(1,2)}(k^2)} - \sqrt{G^{\,\text{\tiny{bl}}}_{(3,0)}}\,, \notag \\[1ex]
\Delta  G_{(1,2)}(0) &=0\,,
\label{eq:inter12}
\end{align}
\end{subequations}
and $\Delta G_{(1,2)}(p^2)$ interpolates between these two values in
the interval $0\leq p^2\lesssim 0.3 \,k^2$, see
\autoref{fig:MomDep}. The accurate determination of this
interpolation at small momenta $p^2 \lesssim 0.3 \,k^2$ is numerically irrelevant
as these momenta are suppressed in loops due to the $p^3$-factor from
the integration measure.

We can make maximal use of the numerical irrelevance of the 
low-momentum regime with $p^2\lesssim 0.3\, k^2$ and drop the non-linear
piece altogether. This amounts to
\begin{align} 
\sqrt{\Delta  G_{(1,2)}(p^2)} &=
 \sqrt{G_{(1,2)}(k^2)} - \sqrt{G^{\,\text{\tiny{bl}}}_{(3,0)}}\,,
\label{eq:inter12-crude}
\end{align}
see also \autoref{fig:MomDep}. In this approximation of the vertex
$p^2\sqrt{G_{(1,2)}(p^2)}$ does not vanish at $p^2=0$, which breaks
shift symmetry. However, the approximation scheme never uses this
information effectively restoring shift symmetry. 

\subsubsection{Qualitative bilocal schemes}
\label{sec:effuni-qual}
A simpler approximation is dropping $\Delta G_{(1,2)}$
completely, $\Delta G_{(1,2)}\equiv 0$. With \eqref{eq:offset12G} this
leads to
\begin{align}\label{eq:DeltaG0}
{G^{\,\text{\tiny{qual}}}_{(1,2)}}=G_{(3,0)}^{\,\text{\tiny{bl}}}\,.
\end{align}
This leads to explicit shift symmetry in \eqref{eq:offset12} but also triggers up to a
$\sim 20\%$ deviation in the results of the respective diagrams proportional to $G_{(1,2)}$.
This approximation has been used with $\Delta G_{(n,m)}=0$ for all $n,m$
in \eqref{eq:offset12} in matter-gravity systems in
\cite{Meibohm:2015twa,Meibohm:2016mkp,Christiansen:2017cxa}, and the
results there receive now support by effective
universality in the scalar-gravity system. 

The final variant of the bilocal scheme is the standard bilocal
approximation for $G_{(1,2)}$. Using shift symmetry with 
$p^2\sqrt{G_{(1,2)}(p^2)}\Big|_{p^2=0}=0$, we are led to 
\begin{align}\label{eq:standard-bi}
G^{\,\text{\tiny{bl}}}_{(1,2)}= G_{(1,2)}(k^2)\,,
\end{align}
for the respective coupling see the right panel of
\autoref{fig:MomDep}. It is up to $\sim 20\%$ bigger than
$G_{(1,2)}(p^2)$ in the numerically relevant regime with
$p^2\gtrsim 0.3\, k^2$. Accordingly, it has a quantitative error of
about this size but maintains explicit shift symmetry. Note also that
it is $\sim 20\%$ smaller than the \emph{slope} of
$p^2\sqrt{G_{(1,2)}(p^2)}$ for $p^2\gtrsim 0.3 \,k^2$ where effective
universality takes place.  This is the approximation we used for the
fixed-point results in \autoref{fig:FPvalues} and also use later in
\autoref{sec:uni-measure}. Based on these observations we call a
deviation from effective universality of up to 20\% a
semi-quantitative agreement of the $\beta$-functions.

\subsubsection{Derivative expansions}
\label{sec:der-exp}
A derivative expansion is a local expansion in momenta. The expansion
point is either chosen for analytic and numerical convenience or in
order to optimise the convergence to the full result. Analytic
convenience singles out $p^2=0$ as this allows for analytic flow
equations for specific regulators such as the Litim regulator
\cite{Litim:2000ci,Litim:2001up} or the sharp cutoff.

Good convergence is usually achieved with an expansion about 
the momentum value where the integrands in the flow peak. 
This typically is a momentum
close to the cutoff scale, $p^2 \approx k^2$, leading to 
\begin{align}\label{eq:derk}
\sqrt{G^{\,\text{\tiny der-k}}_{(1,2)}}= \sqrt{G_{(1,2)}(k^2)}+
\left. p^2\,\0{\partial \sqrt{G_{(1,2)}(p^2)}}{\partial p^2}\right|_{p^2=k^2} \,.
\end{align}
In the present case this
has the additional benefit that it also includes a good estimate of
the linear piece of the $G_{(1,2)}$ avatar of the Newton coupling, see
the right panel of \autoref{fig:MomDep}. Qualitatively is is in the same
ballpark as the quantitative bilocal approximation with
\eqref{eq:inter12-crude} described in \autoref{sec:effuni-quant}.

It is left to discuss the standard derivative expansion with the
expansion point $p^2=0$ with 
\begin{align}\label{eq:der0}
G^{\,\text{\tiny der-0}}_{(1,2)}= G_{(1,2)}(0)\,.
\end{align}
We note that the non-trivial momentum
dependence of the vertex $p^2\sqrt{G_{(1,2)}}$ at small momenta
$p^2 \to 0$ casts some doubt on the naive use of derivative
expansions in quantum gravity. Moreover, the analysis of the 
momentum-space expansion about $p^2=0$ also applies to curvature expansions as
used in the background-field approximation. Hence the current reliability
discussion translates to computations within the 
background-field approximation. 

Clearly, approximating $G_{(1,2)}$ by the derivative of
$p^2\sqrt{G_{(1,2)}(p^2)}$ at $p^2=0$ leads to a significant deviation (factor
$\sim 20$) of the resulting coupling
from the full result in the numerically relevant regime for
$p^2 \gtrsim 0.3\,k^2$. This issue has been already discussed in
\cite{Christiansen:2015rva,Denz:2016qks} for pure quantum gravity,
where the deviation is smaller. This is already visible from the full
momentum-dependence of $G_{(3,0)}$ in \autoref{fig:MomDep}.  Still,
this scheme captures all qualitative aspects of the current system.

\begin{figure*}[t!]
\includegraphics[width=\linewidth]{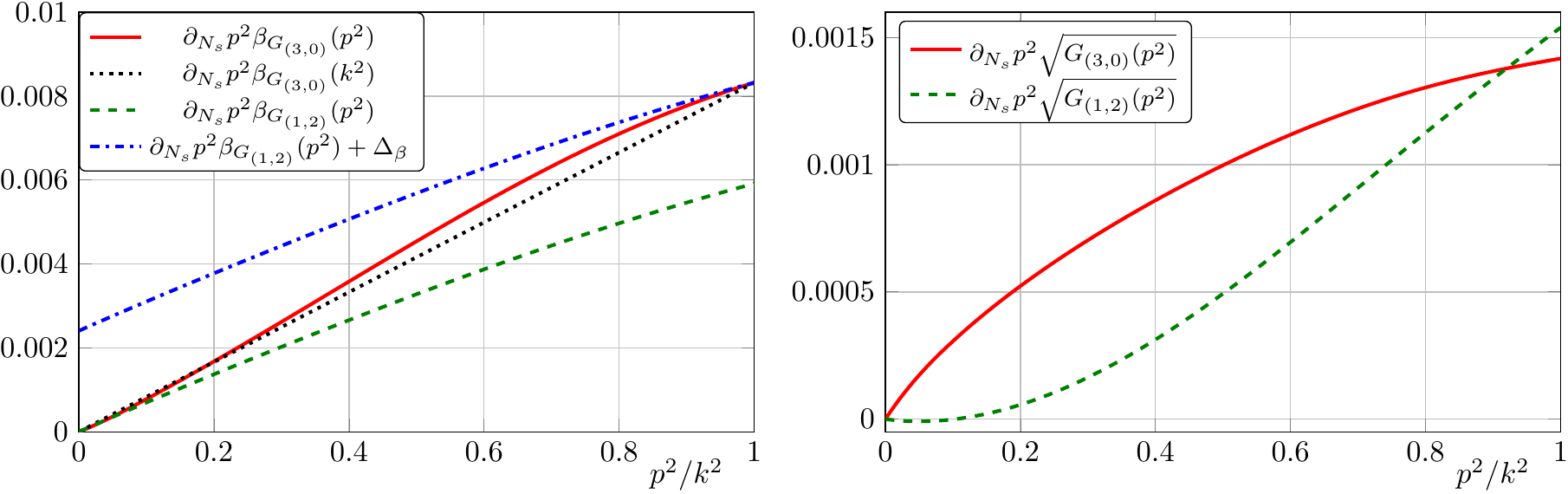}
\caption{$N_s$-derivative of the momentum-dependent
  $\beta$-functions (left) and Newton couplings (right)
  evaluated on the bilocal fixed point for $N_s=0$.}
\label{fig:MomDepNs}
\end{figure*}

\subsection{Effective universality for unquenched quantum gravity}
\label{sec:unquenched}
The quantitative self-consistency analysis in \autoref{sec:quenched}
above was done at $N_s=0$.  Now we study the $N_s$-derivative at
$N_s=0$. This gives us a sum of the different terms that show up with a 
linear $N_s$-dependence that comes from closed scalar loops. If this
sum again shows a behaviour as seen in \autoref{fig:MomDep}, this
indicates the persistence of quantitative effective universality for
all $N_s$. An analysis for large $N_s$ is hampered by the
strongly rising Newton couplings: we hit the reliability bounds of the
approximation before the $N_s$-effects become dominant. 
Thus our analysis is sufficient at the current level of the approximation.

We first examine the $N_s$-derivative of the $\beta$-functions
of the Newton couplings evaluated at the fixed point and then study
the $N_s$-dependence of the momentum-dependent Newton couplings.
For the $N_s$-derivative of the $\beta$-functions we only take the
explicit $N_s$-dependence into account, whereas the $N_s$-dependence of the 
fixed-point values is neglected. They are given by 
\begin{align}
  \partial_{N_s} &p^2 \beta_{G_{(3,0)}}(p^2) 
  \simeq 3 \,G\, p^2\,\partial_{N_s} \eta_h(p^2) \notag\\[1ex]
  &+ (32\pi)^2 \0{64}{171} \sqrt{G} \,\partial_{N_s}\left[ \Flow^{(3,0)}(p^2)-\Flow^{(3,0)}(0) \right] \,,
  \notag \\[1ex] 
  \partial_{N_s} &p^2 \beta_{G_{(1,2)}}(p^2)\simeq
   G\,p^2 \,\partial_{N_s} \eta_h(p^2) \notag\\[1ex]
  &\hspace{2cm}+ \083\sqrt{G} \,\partial_{N_s}\Flow^{(1,2)}(p^2)\,. 
  \label{eq:Nsderbeta}
\end{align}
This  is displayed in the left panel of \autoref{fig:MomDepNs}. We see a qualitatively similar
momentum-dependence of both, $\partial_{N_s} \beta_{G_{(3,0)}}(p^2)$ and
$\partial_{N_s} \beta_{G_{(1,2)}}(p^2)$ evaluated at the fixed
point. Interestingly, the momentum slope of the two $\beta$-functions is almost 
identical at $p^2=0$ and at $p^2=k^2$. Moreover, their absolute values are one/two orders of magnitude
smaller than the quenched $\beta$-functions, see \autoref{fig:MomDep}.
Accordingly, scalars only change the system qualitatively for
$N_s \gtrsim 10 -10^2$. This is seen in \autoref{fig:FPvalues}: the
system is basically unchanged for $N_s\lesssim 20$. From there on the
approximation violates reliability bounds for the regulator with
$\eta_h\leq 2$, see \cite{Meibohm:2015twa}, and should be taken with a
grain of salt.
In general we observe that the $\beta$-functions increase 
with increasing number of scalars,
and hence also the fixed-point values increase with $N_s$.

The $N_s$-derivative of the fixed-point dressing of the vertices for the three-graviton 
coupling read
\begin{subequations}\label{eq:Nsder}
\begin{align}
  \0{\partial_{N_s} p^2\sqrt{G_{(3,0)}(p^2)}}{p^2 \sqrt{G_{(3,0)}(p^2)}} 
  ={}& \0{\partial_{N_s}\left[ \Flow^{(3,0)}(p^2)-\Flow^{(3,0)}(0) \right]}
	{\Flow^{(3,0)}(p^2)-\Flow^{(3,0)}(0)} \notag \\[1ex] 
  &- 3 \0{ \partial_{N_s} \eta_h(p^2)}{2+ 3 \eta_h(p^2) }\,,
  \label{eq:Nsder30}
\end{align}
and for the minimal scalar-graviton coupling 
\begin{align} 
  \0{\partial_{N_s} p^2 \sqrt{G_{(1,2)}(p^2)}}{ p^2 \sqrt{G_{(1,2)}(p^2)} } &=
  \0{\partial_{N_s}\Flow^{(1,2)}(p^2)}{\Flow^{(1,2)}(p^2)} 
  - \0{\partial_{N_s} \eta_h(p^2) }{2+ \eta_h(p^2)  }\,, 
  \label{eq:Nsder12}
\end{align}
\end{subequations}
where the second terms on the right-hand sides of
\eqref{eq:Nsder} take care of the $Z^{n/2}$
dressing of the $n$-point vertices.  The terms in the respective flow
contributions from $\partial_{N_s} \Flow^{(n,m)}$ read
\begin{align}
  \partial_{N_s}\Flow &= \left.\0{\partial }{\partial N_s} \right|_{G,\eta_h}\Flow  
  +[\partial_{N_s}\eta_h(k^2)]\,\partial_{\eta_h} \Flow \notag \\[1ex] 
  &\quad +\032 \0{\partial_{N_s} G}{G} \Flow \,.
  \label{eq:flowNs} 
\end{align}
The first term on the right-hand side of \eqref{eq:flowNs} simply counts
the number of scalars in closed scalar loops, which rises linearly with
$N_s$. This term vanishes for $G_{(1,2)}$ as its flow has no diagram
with a closed scalar loop. The second term takes into account the
$N_s$-dependence of the graviton propagator as well as that of the
wave-function renormalisations in the vertices. With the present
RG-adjusted graviton regulator that is proportional to $Z_h$, this
dependence is stored solely in the $\eta_h$-dependence of the scale
derivative of the regulator. The anomalous dimension $\eta_h$ has a
linear $N_s$-dependence proportional to the closed scalar loop for the
graviton propagator. Together with the closed scalar loop for
$G_{(3,0)}$ it gives the $N_s$-dependence at one-loop. For universal
couplings such as the gauge couplings in the Standard Model these
terms provide the universal $N_s$-dependence of the couplings. 

The additional terms account for the typical resummations present in
FRG computations: an additional contribution in the
$\partial_{N_s}\eta_h$-derivative takes into account the
$N_s$-dependence of the fixed-point coupling. The third term takes
into account the $N_s$-dependence of the fixed-point coupling from the
prefactor $G^{3/2}$ in all the diagrams. Further terms are present
that take into account the $N_s$-dependence of $\mu$ and $\lambda_3$, which
are dropped in the present analysis as they are approximately
$N_s$-independent at the fixed point, see \autoref{fig:FPvalues}. 

This leads us to the right panel of \autoref{fig:MomDepNs},
which displays the $N_s$-derivative of the fixed-point couplings
$p^2 \sqrt{G_{(3,0)}(p^2)}$ and $p^2 \sqrt{G_{(1,2)}(p^2)}$.
The momentum dependence encodes an intriguing structure. First
of all the quantitative effective universality present in
\autoref{fig:MomDep} is not found. Still, the $N_s$-dependences have
the same size, which explains the similar growth in
\autoref{fig:FPvalues}.  Note that this similarity is even better
for the momentum regime relevant in the loop integrals with
$p^2 \approx k^2$, so fully momentum-dependent or bilocal
approximations take account of this fact. 

The momentum-dependence in the right panel of \autoref{fig:MomDepNs}
is not reflected fully by a linear function in $p^2$. This suggests
that higher-order terms are generated by the unquenching terms, i.e.\
the closed scalar loops. We emphasise that the
momentum-dependence in the right panel of \autoref{fig:MomDepNs} is
a superposition of \autoref{fig:MomDep} and the left panel of
\autoref{fig:MomDepNs}. In the latter figures the momentum slopes
did agree well in the momentum regime $p^2\sim k^2$.  Consequently,
also the absolute difference in the momentum slopes at $p^2\sim k^2$
is small in the right panel of \autoref{fig:MomDepNs}.  However, the
relative difference in the momentum slope is large.  Note that
values of the resummed $N_s$-derivative of the momentum-dependent
fixed-point vertices in the right panel of \autoref{fig:MomDepNs}
are even an order of magnitude smaller than the $N_s$-derivative of
the momentum-dependent $\beta$-functions in the left panel of
\autoref{fig:MomDepNs}.  Thus one might potentially interpret our
result as indicating that the $N_s$-dependence of the
momentum-dependent fixed-point vertices is actually compatible with
zero.

As in the quenched scalar-gravity system we now discuss the relation of the
momentum-dependent vertex function to diffeomorphism-invariant operators
by Taylor expanding about vanishing momentum.
Again, this hinges on the existence of a diffeomorphism-invariant expansion 
in the deep UV, where the regulator is expected to spoil diffeomorphism-invariance.
Nonetheless it is worthwhile to examine the overlap of operators with vertex
functions and to check, which inclusion of operators might restore 
effective universality in the unquenched sector.

The $p^2$-terms in $G_{(3,0)}$ and $G_{(1,2)}$ stand for the curvature scalar
$\sqrt{g} R$ and kinetic term $\sqrt{g} \varphi \Delta_g\varphi$,
respectively. The $p^4$-terms in $G_{(3,0)}$ and $G_{(1,2)}$ relate to
$\sqrt{g}R_{\mu\nu}^2$ and 
$\sqrt{g} R_{\mu\nu}\,\varphi \nabla_\mu\nabla_\nu \varphi$ terms respectively, see,
e.g.~\cite{Christiansen:2016sjn,Eichhorn:2017sok}.  Note also that our projection has no
overlap with $\sqrt{g} R^2$ and $\sqrt{g} R\varphi \Delta_g\varphi$.
The $\sqrt{g} R^2$ term is generated with a coupling 
of the same order of magnitude as $G_{(3,0)}$ already in the 
Einstein-Hilbert truncation, see \cite{Denz:2016qks}, and it is only
our projection that has no overlap with it.  In turn, the
$R_{\mu\nu}^2$ coupling generated by the Einstein-Hilbert truncation
is compatible with zero, see also \autoref{fig:MomDep}.
The present finding suggests that the unquenching effects due to the
closed scalar loops generate corrections of the $G_{(3,0)}$-avatar
of the Newton coupling as well as an $R_{\mu\nu}^2$ coupling of a
comparable size. The generation of the latter is particularly intriguing as
the respective term is not generated in quenched scalar-gravity
systems. Consequently, it might be worthwhile to investigate 
effective universality in the unquenched scalar-gravity-system
in the presence of $\sqrt{g}R_{\mu\nu}^2$ and 
$\sqrt{g} R_{\mu\nu}\,\varphi \nabla_\mu\nabla_\nu \varphi$ operators.

\subsection{Effective universality beyond the fixed point}
\label{sec:uni-measure}
The intriguing result displayed in \autoref{fig:MomDep} has shown that
effective universality holds quantitatively at the fixed point within
the $N_s$-range of validity of the current approximation. As argued in
the beginning of \autoref{sec:uni}, we expect effective universality
to only hold in the vicinity of the fixed point, that is on given
trajectories for $k\to\infty$, and, if present, for all momenta at
$k\to 0$. 

Such a scenario suggests an approximation that utilises effective
universality also for finite cutoffs as the related error disappears
at $k=0$ and $k\to\infty$. Here we investigate the question how it
fares away from the fixed point. For the sake of simplicity we do not
resort to the quantitative bilocal scheme described in
\autoref{sec:effuni-quant}, but to the qualitative bilocal scheme
described in \autoref{sec:effuni-qual} with \eqref{eq:standard-bi}. If
evaluating the $\beta$-functions on \eqref{eq:uniG} and on the fixed
point values of $\mu$ and $\lambda_3$ we obtain
\begin{align}
  \beta_{G_{(3,0)}}\Big|_{\mu^*,\lambda_3^*}&=
    2G - ( 3.4 - 0.013N_s )\, G^2 + \CO(G^3)\,, \notag\\[1ex]
  \beta_{G_{(1,2)}}\Big|_{\mu^*,\lambda_3^*}&= 
    2G - ( 2.7 - 0.0085N_s )\, G^2+ \CO(G^3)\,.
  \label{eq:beta-g-comp}
\end{align}
In \eqref{eq:beta-g-comp} we have used the fixed-point values
of $\mu$ and $\lambda_3$ at $N_s=0$, as they are almost
$N_s$-independent (cf.~\autoref{fig:FPvalues}). The coefficients of
the $\beta$-functions \eqref{eq:beta-g-comp} do not feature an effective
universality. This is caused by the missing offset in \eqref{eq:offset12}
at momenta $p^2\gtrsim 0.3\,k^2$ required for the quantitative
agreement, as explained in \autoref{sec:uni-fp}. 

As our further investigation of effective universality is based on the
$\beta$-functions in \eqref{eq:beta-g-comp} we confirm here that the
qualitative nature of the present approximation scheme discussed in
\autoref{sec:effuni-qual} below \eqref{eq:standard-bi} already explains
all the deviations in \eqref{eq:beta-g-comp}:
The $N_s$-independent terms should be subject to underestimating the
\emph{slope} of $p^2\,G_{(1,2)}^*$ in the regime $p^2\gtrsim 0.3\, k^2$
where effective universality takes place, see \autoref{fig:MomDep}.
Accordingly, this part of $\beta_{G_{(1,2)}}$ should be $\sim 20\%$
smaller than that of $\beta_{G_{(3,0)}}$ and their ratio should be
$\sim 0.8$. From \eqref{eq:beta-g-comp} we obtain $2.7/3.4 = 0.79$.

In the $N_s$-dependent part we did not find full effective 
universality, cf.~\autoref{fig:MomDepNs}. However the coefficients
of these terms are two orders of magnitudes smaller than the pure gravity
terms and thus they do not affect the present discussion within
the $N_s$-range of validity of the current approximation, i.e.\ for 
$N_s<50$.

In summary, \eqref{eq:beta-g-comp} fully reflects the quantitative
universality in the given bilocal approximation precisely by its
semi-quantitative or qualitative pattern. This has to be kept in mind if evaluating 
deviations from effective universality within this approximation. 

\begin{figure}[t!]
\includegraphics[width=\linewidth]{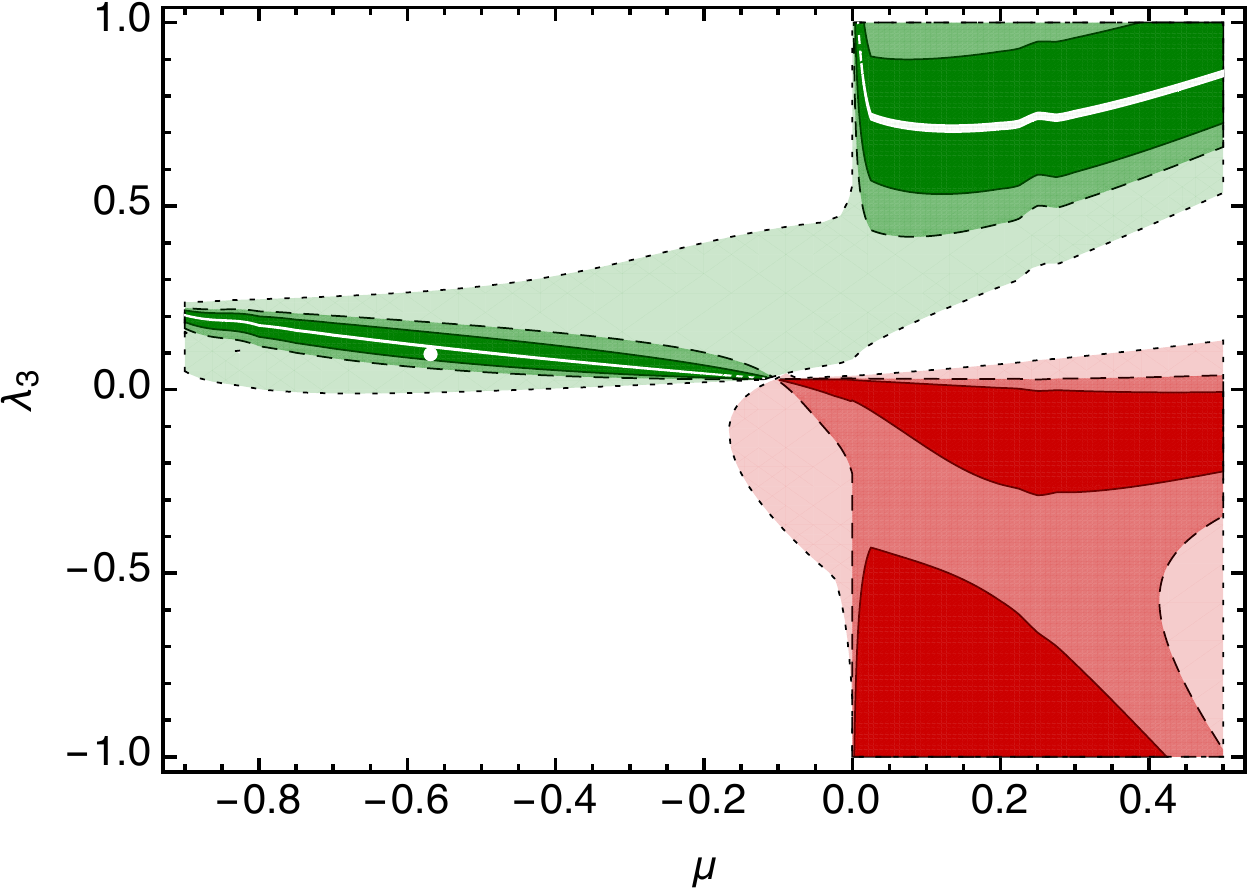}
\caption{
  The regions in which $\varepsilon<\015$ ($\varepsilon<\013$; $\varepsilon<\034$)
  are marked in dark (light, dashed contour; lighter, dotted contour) colours for $N_s=1$ and $G\to0$.
  The green (red) colour indicates $\Delta\beta_{G}<0$ ($\Delta\beta_{G}>0$). 
  The two green regions are centred around a white line where $\varepsilon=0$.
  The bilocal UV fixed point is indicated by a white dot. 
  The area is almost $N_s$ independent, as are the fixed-point values of $\mu$ and $\lambda_3$
  (cf.~\autoref{fig:FPvalues}).}
\label{fig:effunivplane} 
\end{figure}

Now we proceed with this evaluation by devising a measure of the
breaking of effective universality. It is a property of the anomalous 
part $\Delta\beta$ of the $\beta$-function, 
\begin{align}\label{eq:Deltabeta}
 \Delta \beta_{G_{(i,j)}} = \beta_{G_{(i,j)}}-2 G\,.
\end{align}
As a measure of effective universality we use the relative error
between the scaling of two avatars of the coupling evaluated under the
assumption of universality, that is $G_{\vec n}=G$ for the avatars of
the Newton coupling. In the present case this reads
\begin{align}
\varepsilon(G, \mu, \lambda_3, N_s)
&= \left|\frac{\Delta \beta_{G_{(3,0)}} - \Delta \beta_{G_{(1,2)}}}
  {\Delta \beta_{G_{(3,0)}} + \Delta \beta_{G_{(1,2)}}}\right|_{G_{\vec n}=G}\,. 
\label{eq:measureeffu}
\end{align}
In the simplest case of effective universality $\varepsilon$ is
zero. In the present non-trivial realisation we have a breaking
pattern for small momenta, see \autoref{fig:MomDep}. In the presence
of such a breaking $\varepsilon=0$ does indeed indicate a small
violation of effective universality, while a small value might
indicate its full quantitative presence.

Further we observe, that for $\varepsilon<1$ the anomalous parts of
the $\beta$-functions have the same sign, and different signs for
$\varepsilon>1$. In the limit $G\to 0$ we compare the $G^2$-terms as
displayed in \eqref{eq:beta-g-comp}. The definition
\eqref{eq:measureeffu} also allows to separately compare the gravity
and scalar contributions by taking the limits $N_s\to 0$ and
$N_s \to \infty$, respectively. It does, however, not distinguish
between anomalous parts of a $\beta$-function that allow for a UV
fixed point at positive Newton coupling ($\Delta \beta<0$) and that do
not ($\Delta \beta>0$).

In \autoref{fig:effunivplane} we show the regions in the $(\mu, \lambda_3)$-plane
where effective universality is realised for the coupling 
$G_{(3,0)}$ and $G_{(1,2)}$. In particular we display the regions 
in which $\varepsilon<\015$ and $\varepsilon<\013$ for $N_s=1$ and $G\to0$.
We further distinguish between regions that allow for a UV fixed point
($\Delta\beta <0$, green colour) and regions that do not ($\Delta\beta >0$, 
red colour). We observe that effective universality and a UV fixed point 
is only allowed in two regions: this first is for negative $\mu$ and 
small $\lambda_3\approx 0.1$. In fact our fixed-point values for $\mu$ and $\lambda_3$ 
lie in this region of effective universality. In \autoref{fig:effunivplane} it
is marked by a white dot. The other region is at positive $\mu$ and large $\lambda_3>0.5$.
At positive $\mu$ and negative $\lambda_3$ there is another region that allows
for effective universality but not for a UV fixed point at positive Newton coupling.
\autoref{fig:effunivplane} highlights that the common realisation of effective
universality and a UV fixed point is highly non-trivial.

\begin{figure}[t!]
\includegraphics[width=\linewidth]{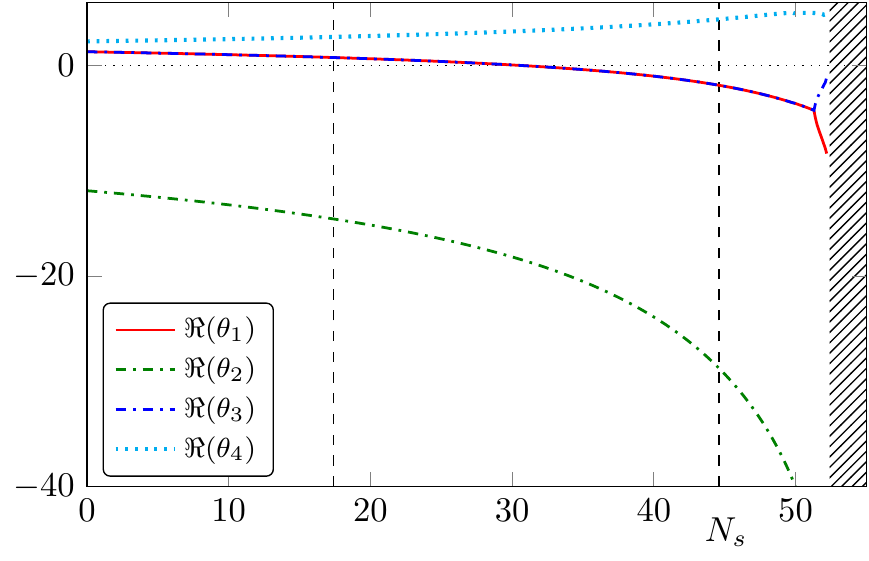}
\caption{Critical exponents of the UV fixed point as a function of $N_s$,
  see \autoref{fig:FPvalues} for the fixed-point values.
  The colours of the critical exponents indicate with which coupling 
  the corresponding eigenvector has the largest overlap.  
  The vertical lines at $N_s \approx 17.5$ and $N_s \approx 44.6$ show where
  $\eta_h(0)$ and $\eta_h(k^2)$ exceed two, respectively.}
\label{fig:allcritexp}
\end{figure}

As effective universality is restored exactly at the fixed point,
i.e.\ if $\mu$ and $\lambda$ are set to their fixed-point values in
$\beta_{G_{(3,0)}}$ and $\beta_{G_{(1,2)}}$, the critical exponents do
not reflect effective universality: the real parts of the relevant
critical exponents actually feature distinct dependencies on $N_s$,
cf.~\autoref{fig:allcritexp}. In particular, the eigendirection which has
most overlap with $G_{(1,2)}$ increases in relevance for increasing
$N_s$.  On the other hand, two superpositions of $G_{(3,0)}$ and $\mu$
form two relevant eigendirections, both of which become less relevant
as $N_s$ is increased. 

\subsection{Effective universality in commonly used approximations}
It is useful to investigate in which commonly used 
approximation effective universality survives.
We have already observed that in our truncation a derivative expansion spoils 
effective universality for the $N_s$-independent part, cf.~\autoref{fig:MomDep},
and it also does not hold anymore away from the fixed point, cf.~\autoref{fig:effunivplane}.
We now take a closer look at the $N_s$-dependent part, where from the left panel of 
\autoref{fig:MomDepNs} we already infer that a derivative expansion works 
rather well.

At $\mu=0=\lambda_3$ and evaluated with a derivative expansion at $p^2=0$, the flow equations for 
the dynamical Newton couplings are given by the analytic expressions
\begin{align}
\beta_{G_{(3,0)}}= (2+ &3 \eta_h ) G_{(3,0)}  - \0{833}{285\pi} G_{(3,0)}^2   \notag \\[1ex]
& - \0{43}{570 \pi} N_s\,G_{(3,0)}^{1/2} G_{(1,2)}^{3/2} \,, \notag\\[1ex]
\beta_{G_{(1,2)}}=  (2 +&\eta_h+ 2\eta_\varphi )G_{(1,2)} \notag\\[1ex]
& - \0{4}{\pi}  G_{(1,2)}^2 + \0{8}{3\pi} G_{(3,0)}^{1/2}G_{(1,2)}^{3/2} \,,
\label{eq:betaG-analytic-fluc}
\end{align}
where we again identified $G_{(n\geq3,0)}= G_{(3,0)}$ and $G_{(n,m\geq2)}=G_{(1,2)}$
and set the anomalous dimension on the right-hand side of the Wetterich equation to zero.
The graviton anomalous dimension in the canonical term gives 
a leading-order contribution and thus we need to include its $N_s$-dependence. With a projection 
at $p^2=0$ the contribution is
\begin{align}
  \eta_h\Big|_{N_s} = \0{1}{24 \pi} N_s\,G_{(1,2)}  \,.
  \label{eq:ns-etah}
\end{align}
A projection at finite momentum changes the size of the contribution
but not the sign. The scalar anomalous dimension does not have any 
$N_s$-dependence.
We use this together with \eqref{eq:betaG-analytic-fluc} and find 
\begin{align}
  \beta_{G_{(3,0)}}\Big|_{N_s}& \approx 0.016\, G^2 N_s \,, \notag\\[1ex]
  \beta_{G_{(1,2)}} \Big|_{N_s}& \approx 0.013 \,G^2 N_s \,.
  \label{eq:Ns-to-fluc-G}
\end{align}
Note that all the above $N_s$-dependent contributions are gauge
independent as they originate from scalar loops.  We find a good
agreement between the two fluctuation couplings
$\beta_{G_{(3,0)}}|_{N_s}$ and $\beta_{G_{(1,2)}}|_{N_s}$, having only
a 16\% deviation of their coefficients.

This computation emphasises that the scalar contribution to different
avatars of fluctuation Newton couplings features a semi-qualitative
agreement already in simple truncations.  This is remarkable, as the
terms from the anomalous dimension $\eta_h$ differ by a factor three
and the agreement originates in a subtraction of the anomalous
dimension and the vertex flow in $\beta_{G_{(3,0)}}$. Moreover, the
behaviour is in agreement with that of $G_{(1,2)}$ in
\cite{Dona:2015tnf}, where a different choice of parameterisation and
gauge was employed. This robustness with respect to (unphysical)
variations of the scheme is encouraging.

\section{Effective universality for the background-fluctuation system}
We now address our second key question and explore whether effective 
universality is also present at the level of the background Newton coupling.
We first investigate this in the full system where the fluctuation couplings 
drive the flow of the background couplings and then we turn to the background 
field approximation and hybrid fluctuation-background--computations.

\begin{figure*}[t!]
\includegraphics[width=\linewidth]{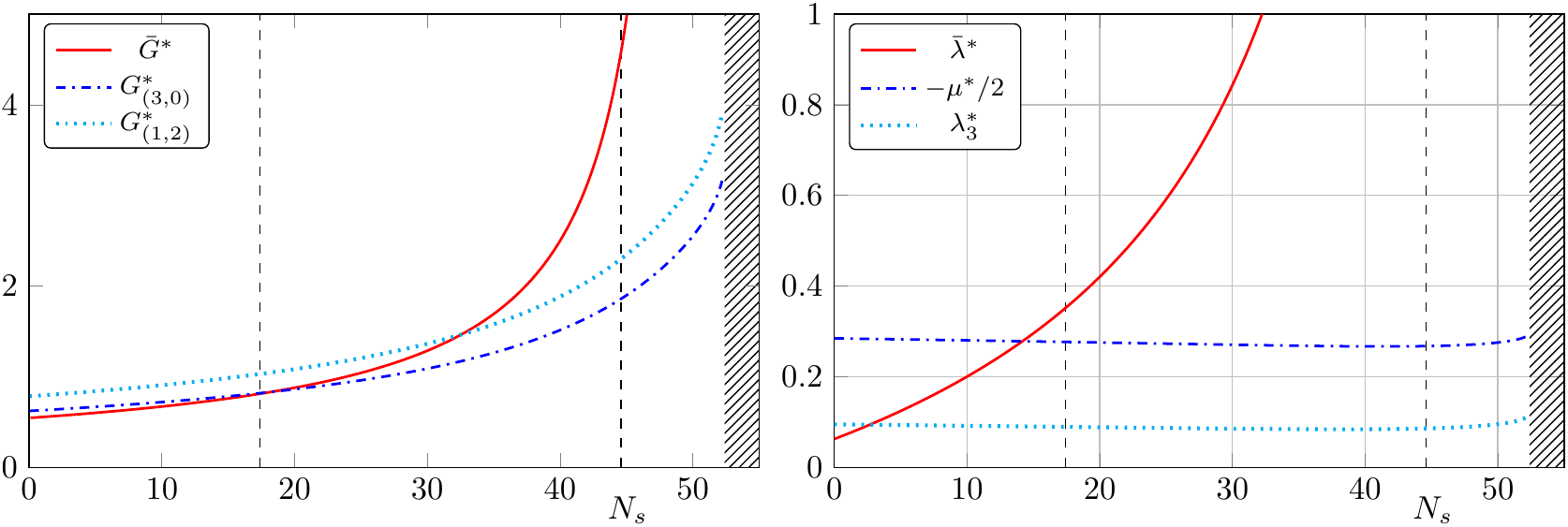}
\caption{Displayed are the fixed-point values as a function of $N_s$.
  We use the fluctuation system as input on the right-hand side of the 
  Wetterich equation.
  Left: Fixed-point values of background and fluctuation Newton couplings.
  Right: Fixed-point values of background and fluctuation cosmological constant.
  The vertical lines at $N_s \approx 17.5$ and $N_s \approx 44.6$ show where
  $\eta_h(0)$ and $\eta_h(k^2)$ exceed two, respectively.
}
\label{fig:Backgr-from-fluc}
\end{figure*}

\subsection{Effective universality for the background Newton coupling}
The flow of $\bar{G}$ and $\bar{\lambda}$ is driven exclusively by the fluctuation 
couplings, their flow equations can be found in App.~\ref{app:background}.
Thus we simply insert the fixed-point value for $\mu$ and the  
fluctuation-field anomalous dimensions on the right-hand side of the Wetterich 
equation, evaluated at $G_{(3,0)}^*, G_{(1,2)}^*, \mu^*$ and $\lambda_3^*$. 

As a function of $N_s$, we observe fixed-point values for $\bar{G}^*$ that track those of the fluctuation
system at the qualitative level, cf.~the left panel of \autoref{fig:Backgr-from-fluc}. A similar 
conclusion can be drawn from the one-loop $\beta$-function for $\bar{G}$, which reads
\begin{align}
\beta_{\bar{G}}= 2\bar{G} - \left(3.64 -0.057\, N_s \right)\bar{G}^2\,. 
\end{align}
Here we have evaluated the $\bar{G}^2$ coefficient on 
$\mu^*$, $\eta_h^* (p^2=k^2)$ and $\eta_c^*(p^2=k^2)$ at $N_s=1$, 
thereby neglecting an additional $N_s$-dependence from the fixed-point couplings themselves.
The pure-gravity coefficient differs by 6\% in comparison to $\beta_{G_{(3,0)}}$,
cf.~\eqref{eq:beta-g-comp}, and the $N_s$-dependent coefficient by a factor $4.4$.
Both signs agree with those in the fluctuation system. The substantial deviation 
of the $N_s$-dependent coefficient leads to a larger gap between the fluctuation 
and the background avatars at large $N_s$. Nevertheless, 
the $N_s$-dependent fixed-point values for all avatars of the Newton coupling agree on the qualitative level.

The fixed-point value of the background coupling $\bar \lambda$ differs
significantly from its fluctuation version, cf.\ the right panel of 
\autoref{fig:Backgr-from-fluc}. Importantly, $\bar \lambda$ and $\lambda_3$ 
can cross the value $\012$, while the graviton mass parameter $\mu$ cannot 
cross the pole located at $\mu=-2 \lambda_2=-1$. As already discussed in 
the beginning of \autoref{sec:uni}, we do not expect effective universality
for those couplings due to the special r\^ole of $\mu$. This again emphasises 
that at least $\mu$ should be taken from a fluctuation computation in order
to obtain robust results. 

\subsection{The fate of effective universality in commonly used approximations}

\subsubsection{Background field approximation}
While it is the fluctuation-field propagator that drives the flow, we
are ultimately interested in the background effective action
$\Gamma_{k\rightarrow 0}[\bar{g}_{\mu\nu}=g_{\mu\nu}, h_{\mu\nu}=0]$
to read off the physics. A commonly used approximation is thus the background 
field approximation, which consists in inserting 
$\Gamma^{(2,0)}_k[\bar{g}_{\mu\nu}=g_{\mu\nu}, h_{\mu\nu}=0]$ on the
right-hand side of the Wetterich equation and thereby letting the
background couplings drive the flow, see also \eqref{eq:gaugesplit} 
and the next section for details. Although this leads to a semi-quantitative agreement with the
full results at $N_s=0$, the approximation fails to capture even the
qualitative $N_s$ dependence correctly. While this might in principle
improve in extended approximations, it casts some doubt on the use of
the background-field approximation for gravity-matter systems at least
in the case of scalar matter.  

For the background Newton coupling, the flow 
equation in the background-field approximation, i.e., at $\eta_h=-2$, evaluated at 
$\bar{\lambda}=0$ reads
\begin{align}
  \beta_{\bar{G}} &=2 \bar{G} 
  - \left(\frac{79}{4}-N_s \right)\frac{\bar{G^2}}{6\pi} \notag\\[1ex]
  &\approx 2 \bar{G} - \left(1.05-0.053N_s \right)\bar{G^2}, 
\end{align}
where the signs of the coefficients still agree with those of the fluctuation system.
The failure of the background-field approximation to correctly capture the $N_s$ dependence 
in our truncation is a consequence of the difference between $\mu$ and $\bar{\lambda}$. While $\mu^*$ 
stays approximately constant with increasing $N_s$, $\bar{\lambda}^*$ is driven towards 
larger values, thereby enhancing gravity fluctuations and suppressing the effect of
scalar-matter fluctuations. The background cosmological constant with fluctuation input 
is displayed in the right panel of \autoref{fig:Backgr-from-fluc}.
The growth of $\bar\lambda^*$ with increasing $N_s$ remains the same in 
the background-field approximation, but it approaches $\bar\lambda^*=\012$
asymptotically, as it cannot cross this pole in this approximation.

\subsubsection{Hybrid scheme for \texorpdfstring{$\eta_h$}{eta h}}
In a hybrid scheme, put forward in \cite{Christiansen:2012rx,Codello:2013fpa} and
employed in the analysis of gravity-matter systems in \cite{Dona:2013qba},
the graviton anomalous dimension is distinguished from the anomalous dimension 
of the background Newton coupling. While $\eta_h$ is evaluated as a function 
of the background couplings $\bar{G}$ and $\bar{\lambda}$ in this hybrid, it can 
deviate from the background-value $\eta_h=-2$, and thereby partially account for 
the nontrivial anomalous dimension of the graviton.
In particular, $\eta_h$ matches that of a fluctuation computation as an expression
of $G$ and $\lambda$; the difference arises from the insertion of different fixed-point
values.

\begin{figure}[t!]
\includegraphics[width=\linewidth]{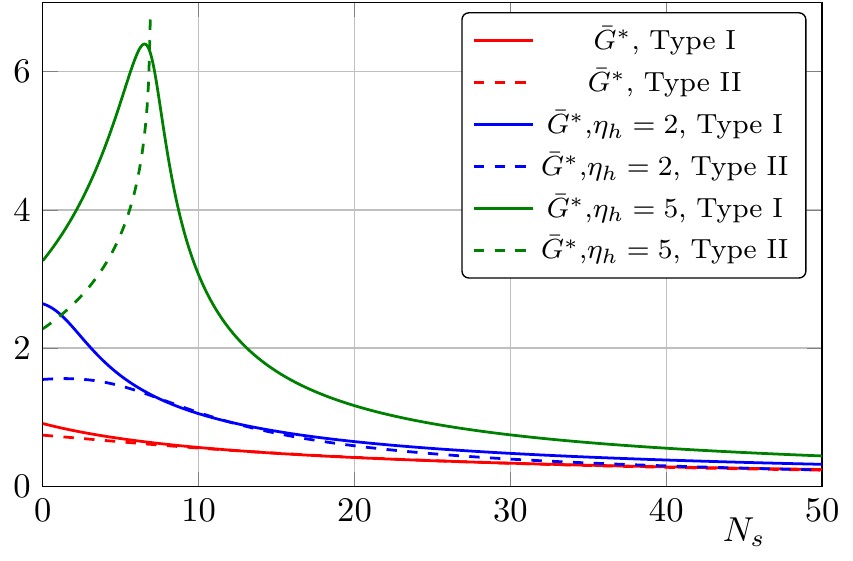}
\caption{Fixed-point values of the background Newton coupling for a
  type-I (continuous lines) and a type-II regulator (dashed lines)
  with the background-field approximation ($\eta_h=-2$, red lines),
  and in hybrid cases with $\eta_h=2$ (blue lines) and $\eta_h=5$
  (green lines). See App.~\ref{app:background} for the definitions of
  type-I and -II regulators.}
  \label{fig:backgroundG_regtype}
\end{figure}

Within such a hybrid setup, a behaviour qualitatively closer to that
of the full fluctuation system was observed, \cite{Dona:2013qba},
i.e.\ the fixed-point value for Newton coupling rose as a function of
$N_s$.  This can be traced back to a growth of the anomalous
dimension, cf.~\autoref{fig:backgroundG_regtype}.  A strong growth of
the anomalous dimension has to be considered carefully: the usual
choice of regulators is $R_k \sim Z_h$, implying a bound on the
anomalous dimensions $\eta<2$ (for bosonic fields)
\cite{Meibohm:2015twa}. As $Z_h \sim k^{-\eta_h}$ in the fixed-point
regime, $\eta_h>2$ destroys the UV behaviour of the regulator that
should suppress all modes in the limit $k \rightarrow \infty$.  For
$\eta_h>4$ signs of diagrams in $\beta$-functions start to flip.
Furthermore a large anomalous dimension can be interpreted as a hint
at large relative cutoff scales between the different fields of the
theory \cite{Christiansen:2017cxa}.

We demonstrate the transition between the $N_s$ dependence of
$\bar{G}^{\ast}$ in the strict background-field approximation and the
hybrid scheme by setting the anomalous dimension to fixed successively
increasing values, cf.~\autoref{fig:backgroundG_regtype}.  We
investigate type-I and type-II regulators \cite{Codello:2008vh}, see
also App.~\ref{app:background} for their definitions.  Indeed the
background Newton coupling with a type-II regulator rises as soon as
$\eta_h>4$ and the diagrams in the $\beta$-function have flipped its
sign.  For the type-I regulator this happens at $\eta_h>6$, as it
features higher powers of $(1-2\lambda)$ in the denominator, which
flip their sign only at $\eta_h>6$.  The fixed-point results for the
Newton coupling in the hybrid scheme therefore match qualitatively
with the effectively universal results of the fluctuation system.

\section{Level-one improvement}
\label{sec:level1}
In \autoref{sec:uni} we found effective universality in
the (quenched) fluctuation system. This effective universality is not present 
in the background system, which only shows a qualitative, not a quantitative 
agreement. This motivates us to investigate at which order of the 
fluctuation computation effective universality sets in. In particular, 
in \eqref{eq:backsplit} we have presented a split of the scale dependent 
effective action in a diffeomorphism-invariant part and a gauge part.
Effective universality hints at a form of the effective action such that
\begin{align}
\Gamma_k[\bar g_{\mu\nu}, h_{\mu\nu},\varphi]={}& 
\Gamma_{k}^{\diff}[g_{\mu\nu},\varphi]
+ \Delta \Gamma_{k}^{\gauge}[\bar{g}_{\mu\nu}] \notag\\[1ex]
&+ h_{\rho\sigma}\Delta \Gamma_{k,\rho\sigma}^{\gauge}[\bar{g}_{\mu\nu}]\,,
\label{eq:Gamma-gauge-lin}
\end{align}
i.e.~an effective action where the gauge part is only linear in the 
fluctuation field. The higher orders are sub-leading.

In this section we upgrade the background couplings to level-one couplings 
with the use of Nielsen identities. By comparing the level-one couplings
to fluctuation ones, we specifically test the importance of 
the term linear in $h_{\mu\nu}$ in \eqref{eq:Gamma-gauge-lin},
i.e.~$\Delta \Gamma_{k,\rho\sigma}^{\gauge}$.

Furthermore, it is desirable to find simple approximations of the system 
that ideally do not require the separate calculation of both background and fluctuation flows.
Therefore we study whether the background-field approximation,
upgraded to a level-one system by using the modified split Ward identity,
can qualitatively or even quantitatively reproduce the behaviour of the full system.

\subsection{Nielsen or split Ward identity and its applications}
\label{sec:Nielsen-id}
We exploit the Nielsen identity (NI) or split Ward identity (sWI) to
improve upon the background-field approximation. Related derivations
and applications in the present context can be found in
\cite{Litim:2002ce,Litim:2002hj,Pawlowski:2003sk,PawlowskiH,Pawlowski:2005xe}.
Split Ward identities in the context of quantum gravity have also been discussed in
\cite{Reuter:2008wj, Reuter:2008qx,
  Donkin,Donkin:2012ud, Becker:2014qya, Bridle:2013sra,
  Dietz:2015owa, Labus:2016lkh, Morris:2016spn,
  Percacci:2016arh, Ohta:2017dsq,Nieto:2017ddk, Safari:2015dva},
  see also \cite{Reuter:1997gx}.

With the introduction of a background field,
the effective action becomes a functional of both
the dynamical fluctuation field $\Phi$ and the auxiliary
background field $\bar\Phi$
\begin{align}
\Gamma_k=\Gamma_k[\bar\Phi,\Phi].
\end{align}
In the present case of
gravity coupled to scalar matter the background and fluctuation fields read
\begin{align}
  \bar\Phi=(\bar g_{\mu\nu}, 0,0,\bar \phi)\,,\qquad 
  \Phi= (h_{\mu\nu},c_{\mu},\bar{c}_{\mu},\varphi)\,,
\end{align}
respectively,
where the full metric and scalar fields are given by 
\begin{align}
g_{\mu\nu}=\bar g_{\mu\nu}+h_{\mu\nu}\,,\quad \phi=\bar\phi+\varphi\,. 
\end{align}
In scalar theories there is no need to choose the cutoff to depend on
the background field. Thus the flowing action is only a function of
the full field $\phi=\bar{\phi}+ \varphi$.  For the purpose of
illustration, we introduce a dependence of the cutoff on $\bar{\phi}$
artificially.  The effective action at $k=0$ is a functional of the
full field $\phi=\bar\phi+\varphi$ only.  This is due to the fact that
the classical action has this property,
$S_\cl [\bar\phi,\varphi]=S_\cl [\bar\phi+\varphi]$.  The shift
symmetry is broken by the cutoff term $R_k=R_k[\bar\phi]$.  The
resulting difference in the dependence on the two fields is sourced
only by the cutoff term and expressed by the NI/sWI, 
\cite{Litim:2002ce,Bridle:2013sra}
\begin{align}\label{eq:scalarNIk}
  \0{\delta \Gamma_k}{\delta\bar\phi}
  - \0{\delta \Gamma_k}{\delta\varphi} =
  \012 \Tr \left[ \0{\delta R_k[\bar\phi]}{\delta \bar\phi}
  G_k[\bar\phi,\varphi]\right] \,.
\end{align}
This equation is derived in straight analogy to the flow equation
itself \eqref{eq:flow}, which is reflected in the structural
similarities \cite{Litim:2002ce}.  We again use the shorthand $G_k$
for the regularised propagator of the fluctuation field, see
\eqref{eq:prop}.  For flows towards the IR where the regulator
vanishes, \eqref{eq:scalarNIk} suggests to use the background-field
approximation
\begin{align}\label{eq:backapprox-b}
	\Gamma_k[\bar\phi,\varphi]\approx \Gamma_k[\bar\phi+\varphi,\varphi=0]\,,
\end{align}
which is exact for $k=0$ in the present example of scalar theories.
However, for flows towards the UV
($k\to\infty$) the background-field approximation is spoiled by
power-counting leading terms in the effective action. This is a
consequence of the mass-like nature of the cutoff, which makes it power-counting UV relevant.
In the following, we do not include the scalar
background field into the regulator, thus the right-hand side of
\eqref{eq:scalarNIk} is zero and we focus on the NI for the graviton.

In gravity, and gauge theories in general, the situation is more
complicated.  In this case there are two sources for the background
dependence of the effective action.  In addition to the regulator
term, the second source of background dependence comes from the gauge
fixing sector $S_\gauge = S_\gf + S_\gh $.  Both the gauge fixing term
$S_\gf$ and the ghost term $S_\gh$ have to depend on the background
field if background gauge invariance is demanded.  Thus the motivation
for introducing the background field, namely background gauge
invariance, leads to a genuine background dependence of the effective
action. For our gravity-matter system the NI \eqref{eq:scalarNIk}
turns into
\begin{align}\notag
  \0{\delta \Gamma_k}{\delta\bar g_{\mu\nu}}
  -\0{\delta \Gamma_k}{\delta h_{\mu\nu}} 
  =& \01{2} \Tr\left[\01{\sqrt{\bar g}}
  \0{\delta \sqrt{\bar g} R_k[\bar g]}{\delta \bar g_{\mu\nu}}
  \,G_k[\bar g, h]\right]\\[1ex]
  &+\left\langle \0{\delta S_\gauge [\bar g,h]}{\delta \bar g_{\mu\nu}}
  -\0{\delta S_\gauge [\bar g,h]}{\delta h_{\mu\nu}}\right\rangle \,,
\label{eq:gaugeNIk}
\end{align}
where the regulator $R_k$ is now a matrix in field space.
The second line in \eqref{eq:gaugeNIk} originates from the background-field
dependence of the gauge fixing sector, and it survives in the limit $k\to 0$ 
if the effective action is evaluated off-shell, but vanishes on the solution of the equations of motion
\cite{Reinosa:2014ooa,Christiansen:2017bsy}. 
In the present work we approximate \eqref{eq:gaugeNIk} and use
\begin{align}
  \lim_{k\to \infty} \left(\0{\delta \Gamma_k}{\delta\bar g_{\mu\nu}}
  -\0{\delta\Gamma_k}{\delta h_{\mu\nu}} \right)\simeq 
  \012 \Tr\left[ \0{1}{ \sqrt{\bar g} } 
  \0{\delta\sqrt{\bar g} R_k[\bar g]}{\delta \bar g_{\mu\nu}}
  \, G_k[\bar g, h]\right]\,,
\label{eq:gravappNIk}
\end{align}
where we have dropped the second line with the gauge fixing
contributions. Two arguments underlie our choice to focus on the
cutoff term: First, at the present level of truncation it is
actually possible to effectively subsume changes in the gauge fixing
under changes of the regulator: Specifically we concentrate on momenta
$p^2 \lesssim k^2$ and a given gauge fixing $S^{(2)}_{\gf}(p^2) $. In
this regime we can utilise the generality of the regulator to
effectively re-adjust it
\begin{align}\label{eq:change-gauge} 
  R_k \to R_k - S^{(2)}_\gf (p^2) \, r(p^2/k^2) 
  + S^{(2)}_{\gf,\diff}(p^2) \,r(p^2/k^2) \,,
\end{align}
where $S^{(2)}_{\gf,\diff}$ is a general gauge fixing term. Hence, with
\eqref{eq:change-gauge} we have effectively changed the gauge fixing term
for momenta $p^2\lesssim k^2$. If applying this procedure to the
ghost, it is only possible to change its propagator and the
interaction of the ghost with the background graviton
$\bar g_{\mu\nu}$, but not that with the dynamical graviton
$h_{\mu\nu}$. As the ghost terms do not take a leading r\^ole
in the flows, this is negligible. In the background-field approximation, and
using the standard expansion in powers of the curvature, the above
mapping strictly holds.  In summary, for the study of different gauge
fixing terms in the present approximation it suffices to study the
regulator dependence of the flow for momenta $p^2\lesssim k^2$. 
In this work we refrain from exploiting this freedom in
practice.

Second, if one compares the contributions of the cutoff term
and the gauge fixing sector to the UV flow, a counting argument 
suggests that the cutoff term dominates. This is because it couples to
all fluctuation modes of the graviton, while the contributions of the
gauge fixing sector couple directly only to the longitudinal modes.
Hence, the transverse-traceless approximation, which focuses 
on the spin-2 mode of the graviton, is only affected by the regulator term 
\eqref{eq:gravappNIk}.

\begin{figure*}[t!]
\includegraphics[width=\textwidth]{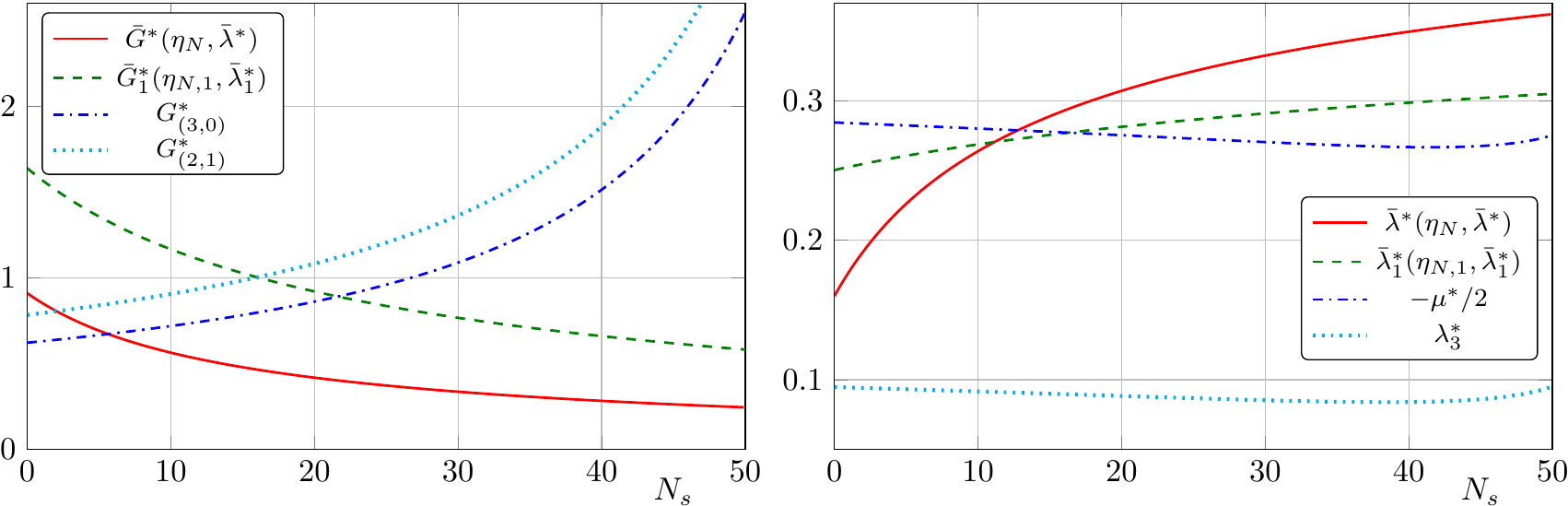}
\caption{Fixed point values of the avatars of the Newton coupling (left panel) 
  and the cosmological constant (right panel) as a function of $N_s$ 
  in the background-field approximation (red continuous lines),
  in the level-one approximation (green dashed lines) 
  and for the fluctuation system (blue dot-dashed and light blue dotted lines).}
\label{fig:level1comp}
\end{figure*}

Finally we are interested in the relation between background and fluctuation field two-point functions. 
To that end we apply $(\delta/\delta {\bar g_{\rho\sigma}} + \delta/\delta {h_{\rho\sigma}})$ to
\eqref{eq:gravappNIk} and drop the cross terms as further approximation, which yields 
\begin{align}\label{eq:gravpropNIk}
  &\0{\delta^2 \Gamma_k}{\delta h_{\rho\sigma}\delta h_{\mu\nu}} 
  -\0{\delta^2 \Gamma_k}{\delta \bar g_{\rho\sigma}\delta \bar g_{\mu\nu}}
  \simeq \\[1ex]
  &\qquad-\012 \left(\0{\delta}{\delta\bar g_{\rho\sigma}}
    +\0{\delta}{\delta h_{\rho\sigma}}\right)
  \Tr\left[\0{1}{\sqrt{\bar g}}\0{\delta \sqrt{\bar g} R_k[\bar g]}{\delta \bar g_{\mu\nu}}\, G_k[\bar
  g, h]\right]\,.\notag 
\end{align}
In the example of Yang-Mills theory \cite{Litim:2002ce,Litim:2002hj},
the analogous fluctuation field derivative of the term in 
the square brackets on the right-hand
side of \eqref{eq:gravpropNIk} gives sub-leading contributions. 
We test a similar assumption and thereby
arrive at the final approximation for the fluctuation two-point function,
which we use in order to close the flow equation,
\begin{align} \label{eq:modback}
  \0{\delta^2 \Gamma_k}{\delta h_{\mu\nu}\delta h_{\rho\sigma}}
  &\approx \0{\delta^2 \Gamma_k}{\delta \bar g_{\mu\nu}\delta \bar g_{\rho\sigma}}  
  - \012\0{\delta}{\delta \bar g_{\rho\sigma}} 
  \Tr\left[ \0{1}{\sqrt{\bar g}}\0{\delta \sqrt{\bar g} R_k}{\delta \bar g_{\mu\nu}}
  \,G_k \right]\,.
\end{align}
This approximation has been used in gravity in~\cite{Donkin,Donkin:2012ud}. 
Apart from the standard background two-point function it contains a second, 
regulator-induced term. Importantly, it can be straightforwardly computed
with heat-kernel methods, see App.~\ref{app:traces} for details.

\begin{figure}[b]
\includegraphics[width=\linewidth]{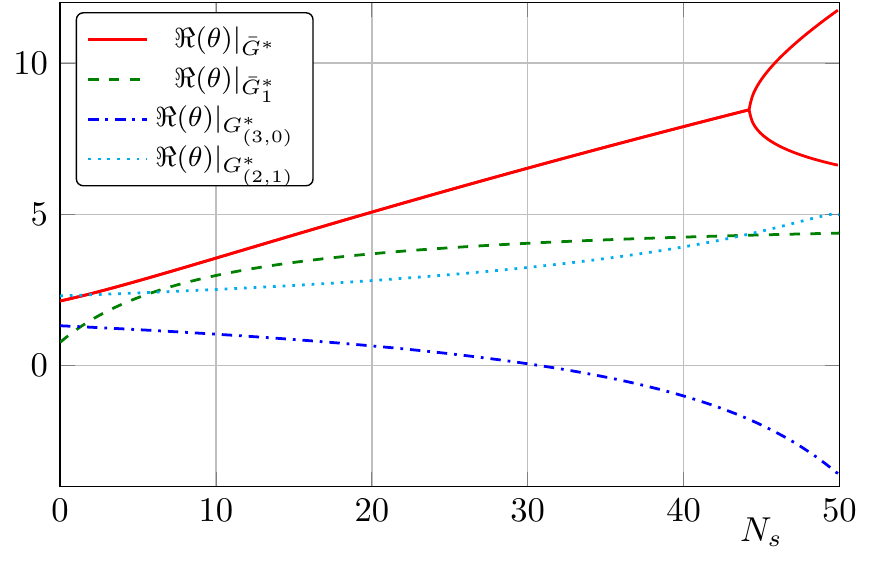}
\caption{Real part of the relevant critical exponents as a
  function of $N_s$ in the background-field approximation (red
  continuous line), in the level-one approximation (green dashed
  line) and for the fluctuation system (blue dot-dashed and light blue dotted lines).
}
\label{fig:critexp}
\end{figure}

\subsection{Fixed-point results for level-one couplings}
Our procedure provides us with a set of $\beta$-functions for the dimensionless 
level-one couplings, $\bar G_1$ and $\bar \lambda_1$, which
are displayed in App.~\ref{app:level-one}.  We now analyse
whether the level-one improvement leads to a system that reproduces the
fluctuation results more closely than the background-field
approximation.

\begin{figure*}[t!]
\includegraphics[width=\textwidth]{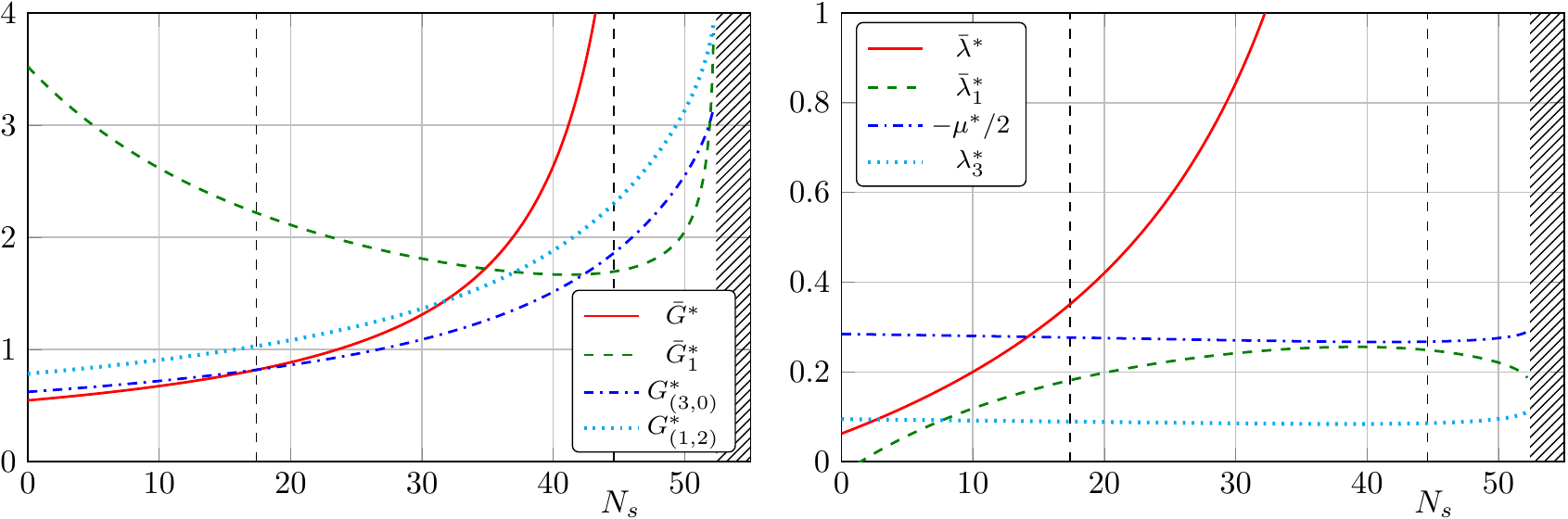}
\caption{Comparison of background, level-one and fluctuation avatars
  of the Newton coupling (left panel) and the cosmological constant (right panel). 
  All couplings are evaluated with the input of the full fluctuation
  system on the right-hand side.
  The vertical lines at $N_s \approx 17.5$ and $N_s \approx 44.6$ show where
  $\eta_h(0)$ and $\eta_h(k^2)$ exceed two, respectively.}
\label{fig:level1fromfluc}
\end{figure*}

In \autoref{fig:level1comp} we display the fixed-point values of the
fluctuation, the level-one, and the background system.  The input on the
right-hand side of the Wetterich equation is fluctuation, level-one, and
background couplings, respectively. In \autoref{fig:critexp} we present the corresponding real
parts of the relevant critical exponents.  The background and the
level-one system each contain exactly two couplings.  Both of them are
relevant and their associated critical exponents form a complex
conjugated pair.  The fluctuation system has four dynamical couplings
and four non-dynamical background couplings (background and level-one
couplings).  Of the four dynamical couplings three are relevant and
one is irrelevant.  Two relevant critical exponents form a complex
conjugated pair and their real part is displayed in
\autoref{fig:critexp} since one can associate them with the couplings
$\mu$ and $G_{(3,0)}$ by means of the largest overlap of the
corresponding eigenvector. In \autoref{fig:level1fromfluc} we
display the fixed-point values of the background, level-one, and
fluctuation couplings with the full fluctuation system as input on the
right-hand side of the Wetterich equation.

We observe that in pure gravity, the level-one improvement leads to
critical exponents that agree better with the fluctuation results than
the background-field approximation, cf.~\autoref{fig:critexp}.  As a
function of $N_s$, the fluctuation results show a qualitatively
different behaviour than the background and the level-one results.  The
real parts of critical exponents of the latter are increasing as a
function of $N_s$ while they are decreasing in the fluctuation case
and even become irrelevant at $N_s\approx 31$.  Nevertheless the
level-one critical exponents are growing slower than the background ones
and thus we observe a slight improvement.

For the fixed-point values of the Newton couplings 
the level-one approximation tracks the background-approximation
in its qualitative dependence on $N_s$: they are both decreasing as a
function of $N_s$, while the fluctuation Newton couplings are
increasing. However the quantitative difference between $G^*_{(3,0)}$
and $\bar G^*_1$ is smaller than the quantitative difference between
$G^*_{(3,0)}$ and $\bar G^*$, cf.~\autoref{fig:level1comp}. In the sector
of the cosmological constants or momentum-independent parts of the
$n$-point functions, the level-one improvement is more pronounced.
While the background cosmological constant increases strongly and
approaches the pole at $\bar \lambda=\012$, the level-one coupling
remains almost constant and increases only slightly.  The fluctuation
coupling $-\mu/2$ also remains almost constant but decreases slightly
with $N_s$.

In summary the level-one approximation might be considered a slight
improvement over the background-field approximation.  We have observed
slight improvements in the critical exponents and in the fixed-point
values of the Newton couplings and the cosmological constants.
Considering however its failure to adequately capture the fluctuation results, 
a level-one approximation seems hardly justified in view of the significantly increased computational
effort---at least based on the results in our truncation.

Last but not least we consider the fixed-point results when
all couplings, including the background and the
level-one coupling, are evaluated with the fluctuation couplings on the right-hand
side of the Wetterich equation, cf.~\autoref{fig:level1fromfluc}. We
observe that the level-one approximation even appears to break the
effective universality that was observed for the Newton coupling: the
qualitative and quantitative dependence of $\bar G_1^*$ on $N_s$ does
not match that of the other couplings as it first decreases with $N_s$
and then increases strongly. For the `cosmological constants', we
make the opposite observation: while the background cosmological
constant deviates strongly from the $N_s$-dependence of the graviton
mass parameter, the level-one cosmological constant approaches it
towards larger $N_s$.  For the canonically most relevant coupling in
the truncation, the step from the background coupling to the level-one
coupling is therefore a significant step towards capturing the behaviour
of the fluctuation coupling.

\section{Summary and outlook}
In this work we have focused on two key questions within the
asymptotic safety program for quantum gravity.

Firstly, we have introduced and explored the concept of \emph{effective
  universality} for the dynamical couplings in matter-gravity
systems. Gauge theories always feature several avatars of the same
coupling, such as, e.g. the quark-gluon coupling and the three-gluon
and four-gluon coupling in QCD. These couplings are related by gauge
invariance. Together with the marginal nature of these couplings in
$d=4$ and the corresponding two-loop universality this offers a
universal definition of the gauge coupling. 

In gravity in $d=4$, the Newton coupling is dimensionful and a similar
form of universality is therefore not to be expected. Different
avatars of the Newton coupling, defined, e.g. from the three-graviton
vertex and the scalar-graviton vertex, are therefore by no means
guaranteed to agree. On the other hand, diffeomorphism invariance of
course implies a relation between those couplings.  We define 
\emph{effective universality} as a quantitative agreement of these
couplings, at least at the asymptotically-safe fixed point. In this work, the
viability of this concept has been explored in a scalar-gravity-system
with $N_s$ scalars. We have specifically focused on the
scalar-graviton coupling and the three-graviton coupling. We have
discovered indications for effective universality in the quenched
approximation, where closed scalar loops are neglected. Apart from a
difference between the two Newton couplings at small momenta, they
show a remarkable quantitative agreement, in particular in the most
relevant range of momenta close to the cutoff. "Unquenching" quantum
gravity leads to $N_s$-dependent corrections which are subleading in
the tentative range of validity of our truncation. While effective
universality in the sense of a quantitative agreement is lost, we
still observe qualitative agreement of the contributions. We have also
discussed effective universality in commonly used approximations,
ranging from bilocal schemes to the derivative expansion. We find
that effective universality even holds for the derivative expansion of
the flows about vanishing momentum, despite the large quantitative
deviations from the full flows. Obtaining quantitative results requires going
beyond the derivative expansion using at least bilocal schemes. 

Our results provide a justification for a commonly used approximation,
in which different avatars of the Newton coupling are equated. In
terms of exploring the consequences of asymptotically safe quantum
gravity within efficient, i.e. small, truncations, effective
universality is key.

Moreover, the emergence of effective universality at the fixed point
is a very strong indication for the physical nature of the
asymptotically-safe fixed point, in the following sense: For a
truncation-induced fixed point, there is no reason to exhibit
effective universality, since it is simply a ``random" solution to a
set of polynomial equations. On the other hand, for an actual fixed
point, diffeomorphism invariance provides a mechanism to restore
effective universality at the fixed point. In particular, in our
study, even the qualitative agreement of the beta functions for
different avatars of the Newton coupling in the unquenched setting
requires non-trivial cancellations between different contributions to
the beta functions. In our opinion these cancellations are extremely
unlikely to randomly occur in truncation-induced fixed points. 

We then have proceeded to tackle the second key question, namely the relation
of the dynamical system to the background system. We have focused on a search
for effective universality, and discovered a qualitative agreement of
the beta function for the background Newton coupling with that of the
dynamical system. However, effective universality is lost, casting
doubt on the use of the background approximation for quantitatively
reliable results. Moreover, the beta functions for the cosmological
constant and graviton mass parameter are manifestly different, leading
to different behaviour in the two systems at large $N_s$ in our
truncation. In practise, our results can be read as tentatively
suggesting that at least the graviton mass parameter should always be
taken from a fluctuation calculation.

Finally, we explore whether an upgrade of the background system to a
level-one system by means of the modified Nielsen or shift Ward identity
restores effective universality. Focusing on the regulator
contribution to the Nielsen identity, we find that the level-one
upgrade is insufficient. Still, interestingly
the level-one cosmological constant is the only coupling that
significantly improves towards the corresponding fluctuation result
for the graviton mass parameter.

In summary, our results yield an a-posteriori justification for the
use of relatively simple truncations that make use of effective
universality. Even more important, the highly nontrivial signatures 
for effective universality in scalar-gravity systems provide further
evidence for asymptotic safety in quantum gravity.

\vspace{.5cm}
\noindent \textbf{Acknowledgements} We thank N.~Christiansen,
S.~Lippoldt and R.~Percacci for discussions.  AE is supported by an
Emmy-Noether-grant of the DFG under AE/1037-1. PL thanks the ITP
Heidelberg for hospitality during a research stay funded by the DFG
grant AE/1037-1.  MR acknowledges funding from IMPRS-PTFS. This work
is supported by EMMI and is part of and supported by the DFG
Collaborative Research Centre "SFB 1225 (ISOQUANT)".

\appendix

\section{Background flow equations}
\label{app:background}
In this Appendix we display the background-flow equations.
We use a Litim-type regulator \cite{Litim:2000ci,Litim:2001up}
for all fields in our setup
\begin{align}\nonumber  
 R^{ij}_k(p^2) &= \delta^{ij}\,
 \left. \Gamma^{(\phi_i \phi^*_i )}(p^2)\right|_{\mu=0} r_{\phi_i}(p^2/k^2)\,,\\
\qquad 
r(x) &=\left(\01x -1\right) \theta(1-x)\,.
\label{eq:flat}
\end{align}
We employ the gauge $\alpha=\beta=0$ and use the Laplacian
as coarse-graining operator, i.e., a type-I regulator.
In \autoref{fig:backgroundG_regtype} we also used a type-II regulator for
comparisons, which means that we used the Laplacian plus explicit curvature 
terms of the two-point functions as coarse graining operator,
see \cite{Codello:2008vh} for more details on the different types of regulators.
We further employ the York-decomposition \cite{York:1973ia,Stelle:1976gc}
for the graviton 
\begin{align}
 h_{\mu\nu}=  h^{\tinyTT}_{\mu\nu} + \frac{1}{d} \bar g_{\mu\nu} h^{\tinytr} 
 + 2 \bar{\nabla}_{(\mu}\xi_{\nu)}+
 \left(\bar{\nabla}_{\mu}\bar{\nabla}_{\nu}-\frac{\bar g_{\mu\nu}}{d}\bar{\nabla}^2\right) \sigma\,,
\end{align}
and the ghost
\begin{align}
 c_\mu = c_\mu^{\text{\tiny T}} + \bar{\nabla}_{\mu} \eta\,,
\end{align}
with field redefinitions 
according to \cite{Dou:1997fg,Lauscher:2001ya,Gies:2015tca}
\begin{align}
 \xi^\mu &\rightarrow  \01{\sqrt{\bar \Delta - \0{\bar R}4}} \xi^\mu\,, \notag\\
 \sigma &\rightarrow \01{\sqrt{\bar \Delta^2 - \bar \Delta\0{\bar R}3}} \sigma\,, \notag\\
 \eta &\rightarrow \01{\bar \Delta} \eta\,.
\end{align}
The result for the gravity part of the
background flows agrees with e.g.~\cite{Gies:2015tca,Denz:2016qks} and the scalar
part agrees with e.g.~\cite{Dona:2013qba}.
The details of the computation are given in App.~\ref{app:traces}.
The flow of the background couplings are given by
\begin{align}
 \partial_t \bar G &= (2 + \eta_N) \bar G \,,\notag\\[1ex]
 \partial_t \bar \lambda &= - 4 \bar \lambda + \0{\bar \lambda}{\bar G} \partial_t \bar G 
	+ 8 \pi \bar G \Flow_{\bar \Gamma}\bigg|_{\sqrt{\bar g} \text{-terms}} \,,\notag\\[1ex]
 \eta_N &= 16 \pi \bar G \Flow_{\bar \Gamma}\bigg|_{\sqrt{\bar g} \bar R \text{-terms}} \,,
\end{align}
where
\begin{widetext}
\begin{align} \label{eq:background-flow}
  \Flow_{\bar \Gamma} = \0{ \sqrt {\bar g} }{ 32 \pi^2 }
  \bigg\lbrace
    &\bigg[
      \0{5 - \0{5}{6} \eta_h}{1 - 2 \lambda}
      + \0{1 - \0{1}{6} \eta_h}{1 - \0 43 \lambda} 
      - 4
      - \0{2}{3} \, \eta_h
      + \0{4}{3} \, \eta_c
      + N_s (1 - \016 \eta_\varphi)
    \bigg] k^4 
     \\[1ex]
    + &\bigg[
      - \0{10}{3} \0{1 - \0{1}{6} \eta_h}{(1 - 2 \lambda)^2}
      - \0{5}{3} \0{1 - \0{1}{4} \eta_h}{1 - 2 \lambda}
      + \0{1}{3} \0{1 - \0{1}{4} \eta_h}{1 - \0 43 \lambda} 
      - \0{23}{12}
      - \0{7}{18} \, \eta_h
      + \0{7}{9} \, \eta_c
      + \0 {1}{3} N_s (1 - \014 \eta_\varphi)
    \bigg] k^2 \bar R 
  \bigg\rbrace
    + \mathcal O (\bar R^2) \,.\notag
\end{align}
\end{widetext}
Note that the quantities $\lambda$, $\eta_h$, $\eta_c$, and $\eta_\varphi$ 
can be taken from the respective fluctuation two-point functions.
In this case the background couplings are non-dynamical spectators.
The usual background-field approximation is obtained
by setting $\lambda=\bar\lambda$, $\eta_h=\eta_N$, $\eta_c=0$, and $\eta_\varphi=0$.

\section{Level-one flow equations}
\label{app:level-one}
In this Appendix we display the level-one flow equations that are derived 
though a Nielsen identity from the background-flow equations, see \autoref{sec:level1}.
We work with the approximation
\begin{align}
 \0{\delta}{\delta h_{\mu\nu}} \partial_t \Gamma_k \approx 
 \0{\delta}{\delta {\bar g}_{\mu\nu}} \partial_t \Gamma_k  
 - \partial_t \left( \012 \Tr\left[\01{\sqrt{\bar g}}
 \0{\delta \sqrt{\bar g} R_k }{ \delta {\bar g}_{\mu\nu} }\, G_k\right] \right)\,.
\end{align}
Consequently we are interested in evaluating 	
\begin{align}
  \CI^{\mu\nu} = I \,\bar{g}^{\mu\nu} =
  \0 12  \Tr\left[\01{\sqrt{\bar g}}\0{\delta \sqrt{\bar g} R_k }{ \delta {\bar g}_{\mu\nu} }\, G_k \right] \,,
  \label{eq:def-I}
\end{align}
which gives us, combined with the background flows, the flows for the level-one couplings.
The trace appearing in \eqref{eq:def-I} can be evaluated using heat kernel techniques.
Details of the computation are presented in App.~\ref{app:traces}.
The result for $I$ is given by
\begin{widetext}
\begin{align} \label{eq:sol-I}
  8I = \0{ \sqrt{\bar g} }{ 32 \pi^2 } \,
  \bigg\lbrace
    &\bigg[
    \0 {10}{3} \0{1}{1 - 2 \lambda}
    + \0 2{3} \0{1}{1 - \0 43 \lambda}
    - \0{8}{3}
    + \0{2}{3}N_s
    \bigg] k^4 \notag\\[1ex]
    +&\bigg[
      - \0 {20}{9} \0{1}{(1 - 2 \lambda)^2}
      - \0 {5}{2} \0{1}{1 - 2 \lambda}
      + \0 {1}{2} \0{1}{1 - \0 43 \lambda}
      - \0 {71}{36}
      + \0 {1}{2} N_s
    \bigg] k^2 \bar R
    \bigg\rbrace
    + \mathcal O (\bar R^2) \,.
\end{align}
\end{widetext}
We display the result for $8I$, since the Nielsen identity
enters precisely with this factor in the flow equation for the $\sqrt{\bar g}$-terms
as well as the $\sqrt{\bar g} \bar R $-terms.
The reason for this is that $\CI^{\mu\nu} $ enters with a $\partial_t$
derivative, $\partial_t \sqrt{\bar g} k^4 = 4 \sqrt{\bar g} k^4$ and 
$\partial_t \sqrt{\bar g} \bar R k^2 = 2 \sqrt{\bar g} \bar R k^2$,
while $\Flow_{\bar \Gamma}$ enters with a $\delta_{\bar g}$ derivative,
$\delta_{\bar g}\sqrt{\bar g} = \012 \sqrt{\bar g}\bar g^{\mu\nu}$
and $\delta_{\bar g}\sqrt{\bar g} \bar R = \014 \sqrt{\bar g}\bar g^{\mu\nu}\bar R$
cf.~\eqref{eq:delta-bar-g}.
Consequently in both cases they combine to the factor 8 as indicated above.
Note that $\CI^{\mu\nu}$ does not contribute to the flow equation 
for $\sqrt{\bar g} \bar R^2$ since $\partial_t \sqrt{\bar g} \bar R^2 k^0 = 0$.
This is expected since $R^2$ and $R_{\mu\nu}^2$ are marginal couplings and 
thus their one-loop flow equations are universal.
Note furthermore that in the above discussion we have neglected terms like 
$\partial_t \lambda$. Such terms do not change the fixed-point values 
but they can influence the critical exponents.

The flow equations for the level-one couplings can now be expressed 
by flow of the background couplings plus the improvement from the Nielsen identity, to wit
\begin{align}
 \partial_t \bar G_1 &= (2 + \eta_{N,1}) \bar G_1 \,, \notag\\[1ex]
 \partial_t \bar \lambda_1 &= - 4 \bar \lambda_1 + \0{\bar \lambda_1}{\bar G_1} \partial_t \bar G_1 
	  + 8 \pi \bar G_1 \left(  \Flow_{\bar \Gamma} - 8 I \right) \bigg|_{\sqrt{\bar g} \text{-terms}} \,,\notag \\[1ex]
 \eta_{N,1} &= 16 \pi \bar G_1 \left(\Flow_{\bar \Gamma} - 8 I  \right)\bigg|_{\sqrt{\bar g} \bar R \text{-terms}} \,.
 \label{eq:level1betas}
\end{align}
Again, the quantities $\lambda$, $\eta_h$, $\eta_c$, and $\eta_\varphi$ 
can be taken from the respective fluctuation two-point functions.
Then the level-one couplings are non-dynamical spectators.
Otherwise we can close the equation at the level-one couplings
by setting $\lambda=\bar\lambda_1$, $\eta_h=\eta_{N,1}$ and $\eta_c=\eta_\varphi=0$. 
The latter is an improved background-field approximation.

\section{Evaluation of traces}
\label{app:traces}
In this Appendix we present the computation of the Nielsen identity from \autoref{sec:Nielsen-id}.
The computation of the trace-term in \eqref{eq:modback} is the challenging part.
For this we first expand the propagator in orders of background curvature
\begin{align}
  \label{eq:expansion-tr}
   \CI^{\mu\nu} &\equiv I\, \bar g^{\mu\nu} := 
   \012 \Tr\left[\01{\sqrt{\bar g}}\0{\delta \sqrt{\bar g} R_k }{ \delta {\bar g}_{\mu\nu} }\, G_k\right]  \notag\\[1ex]
  &= \014 \,\bar g^{\mu\nu} \Tr\left[R_k G_k\right] 
    + \012 \Tr\left[\0{\delta R_k }{ \delta {\bar g}_{\mu\nu} }\, G_k(\bar R = 0)\right] \notag \\[1ex]
    &\quad+ \012 \bar R \, \Tr\left[\0{\delta R_k }{ \delta {\bar g}_{\mu\nu} }\, G'_k(\bar R = 0)\right] 
    +\CO(\bar R^2)\,.
\end{align}
For the terms involving a background derivative of the regulator we use that
\begin{align}
 \0{\delta R_k(\bar \Delta)}{ \delta \bar g_{\mu\nu}} &G_k(\bar \Delta, \bar R = 0) 
 = \0{\delta \bar \Delta}{ \delta \bar g_{\mu\nu}} \0{\partial R_k(\bar \Delta)}{\partial \bar \Delta} G_k(\bar \Delta, \bar R = 0) \notag \\[1ex] 
 &= \0{\delta}{\delta \bar g_{\mu\nu}} \int_{0}^{\bar\Delta} d x' \0{\partial R_k(x')}{\partial x'} G_k(x', \bar R = 0) 
 \,, \label{eq:der-F}
\end{align}
and define the latter integral as
\begin{align}
  \label{eq:Fg}
  F_{\RG}(x)= \int_{0}^x \mathrm dx' \;
  \0{\partial R_k(x')}{\partial x'} \, G_k(x', \bar R = 0)\,,
\end{align}
where we have restricted ourselves to IR-finite regulators.
In straight analogy we manipulate the trace term with $G'$ in 
\eqref{eq:expansion-tr} and define
\begin{align}
  F^{(1)}_{\RG}(x)= \int_{0}^x \mathrm dx' \; \0{\partial R_k(x')}{\partial x'} \, G^{(0,1)}_k(x', \bar R = 0) \,.
\end{align}
As we will see later, these terms only contribute at order $\bar R^2$
and are thus not relevant for the present work.
This leads as to 
\begin{align} \label{eq:def-I-long}
  \CI^{\mu\nu} &=
  \0 14 \, \bar g^{\mu\nu} \, \Tr \left[ R_k G_k \right] 
  + \0 12 \, \Tr \left[\0{\delta}{\delta \bar g_{\mu\nu}} \, F_{\RG} \right] 
  \notag \\[1ex]
  &\quad + \0 12 \, \bar R \, \Tr \left[\0{\delta}{\delta \bar g_{\mu\nu}} \, F^{(1)}_{\RG} \right]
  + \CO(\bar R^2)\,,
\end{align}
where we can pull the $\bar g$-derivative out of the trace.
In this work we have defined the trace such that we have to there is a factor $\sqrt{\bar g}$ involved.
Thus we obtain
\begin{align} \label{eq:expanded-I}
 \CI^{\mu\nu} &= \014 \,\bar g^{\mu\nu} \Tr\left[G_k R_k\right] 
    +\012 \0{\delta}{\delta \bar g_{\mu\nu}} \Tr\left[F_{\RG}\right]
    -\014 \,\bar g^{\mu\nu} \Tr\left[F_{\RG}\right]
    \notag \\[1ex]
    &\quad +\012  \bar R \0{\delta}{\delta \bar g_{\mu\nu}} \Tr\left[F^{(1)}_{\RG}\right] 
    -\014 \,\bar g^{\mu\nu} \bar R\,  \Tr\left[F^{(1)}_{\RG}\right] 
    +\CO(\bar R^2) \,.
\end{align}
We now specify these traces to their contributions in order in the background curvature.
For this we need to evaluate the background-field derivative.
Since $S^4$ is an Einstein manifold, and in particular
\begin{align}
  \bar R_{\mu\nu} = \01d \bar R\, \bar g_{\mu\nu} \,.
\end{align}
Applying a background-field derivative, this gives 
\begin{align}
  \0{\delta \bar R}{\delta \bar g_{\mu\nu}} = 
  - \01d \bar R\, \bar g^{\mu\nu} + \, \bar g^{\alpha\beta} \0{\delta \bar R_{\alpha\beta}}{\delta \bar g_{\mu\nu}} \,.
\end{align}
We further know that
\begin{align}
    \0{\delta \bar R_{\alpha\beta}}{\delta \bar g_{\mu\nu}} 
    = \bar \Delta_\rho \0{\delta \Gamma^\rho_{\alpha\beta}}{\delta \bar g_{\mu\nu}} 
    - \bar \Delta_\alpha \0{\delta \Gamma^\rho_{\rho\beta}}{\delta  \bar g_{\mu\nu}} \,,
\end{align}
are total derivatives and do not contribute.
In summary we have
\begin{align}
  \0{\delta \sqrt{\bar g}}{\delta \bar g_{\mu\nu}} = \012 \sqrt{\bar g}\, \bar g^{\mu\nu} \,,
  \qqquad
  \0{\delta \sqrt{\bar g} \bar R}{\delta \bar g_{\mu\nu}} = \0 14  \sqrt{\bar g}\, \bar g^{\mu\nu} \bar R \,.
  \label{eq:delta-bar-g}
\end{align}
Using this for \eqref{eq:expanded-I} we obtain
\begin{align} \label{eq:I-slim}
 I\bigg|_{\sqrt{\bar g}\text{-terms}} &= 
  \014 \, \Tr\left[G_k R_k\right] \bigg|_{\sqrt{\bar g}\text{-terms}} \,,\notag \\[1ex]
 I\bigg|_{\sqrt{\bar g}\bar R\text{-terms}} &= 
  \014 \left( \Tr\left[G_k R_k\right] 
  -\012 \, \Tr\left[F_{\RG}\right] \right) \bigg|_{\sqrt{\bar g}\bar R\text{-terms}} \,.
\end{align}
Remarkably the $F_\RG$-traces do not contribute to the $\sqrt{\bar g}$-terms 
and the $F^{(1)}_{\RG}$-traces drop out completely.
The latter would contribute to the $\sqrt{\bar g} \bar R^2$-order.
In summary we compute the traces 
$\Tr\left[G_k R_k\right]$, $\Tr\left[F_{\RG}\right]$ 
and $\Tr\left[G_k \partial_t R_k\right]$. 
The latter we need for the flow of the background couplings, 
which are also the basis for the level-one flow equations.

With the gauge choice $\beta = 0$ and $\alpha \rightarrow 0$,
the product of $G_k R_k$ is given by the sum of the contributions
of the four modes $h^{\tinyTT}_{\mu\nu}, \xi^{\mu}, h^\tinytr$, and $\sigma$. 
They are given by
\begin{align} \label{eq:graviton-modes}
(GR)_{h_\tinyTT} ={}& \0{r_k(x)}{x + r_k(x) - 2  \lambda} 
		- \0 23  \0{\bar R }{k^2} \0{r_k(x)}{(x +r_k(x) - 2  \lambda) ^2}
		\notag\\[1ex] 
		&+\mathcal{O}(\bar R^2) \,,\notag\\[1ex]
(GR)_{\xi} ={}& \0{r_k(x)}{x + r_k(x)} 
		+ \0 14 \0{\bar R }{k^2} \0{r_k(x)}{(x + r_k(x)) ^2} 
		+\mathcal{O}(\bar R^2) \,,\notag\\[1ex]
(GR)_{h_\tinytr} ={}& \0{r_k(x)}{x+r_k(x) - \0 43 \lambda} \,,\notag\\[1ex]
(GR)_{\sigma} ={}& \0{r_k(x)}{x+r_k(x)} 
		+ \013  \0{\bar R }{k^2} \0{r_k(x)}{(x+r_k(x)) ^2} +\mathcal{O}(\bar R^2)\,,
\end{align}
where $x=\bar \Delta/k^2$ is the dimensionless background Laplacian.
The vector and scalar ghost modes, $c$ and $\eta$, respectively, are given by
\begin{align} \label{eq:ghost-modes}
(GR)_{c} &=  \0{r_k(x)}{x+r_k(x)} 
	    + \014 \0{\bar R }{k^2} \0{r_k(x)}{(x+r_k(x)) ^2}  + \mathcal{O}(\bar R^2) \,, \notag \\[1ex]
(GR)_{\eta} &=  \0{r_k(x)}{x+r_k(x)} 
	      + \013 \0{\bar R }{k^2} \0{r_k(x)}{(x+r_k(x)) ^2}  + \mathcal{O}(\bar R^2)\,.
\end{align}
The generalisations to $G_k \partial_t R_k$ and $G_k \partial_{\bar \Delta} R_k$
are straightforward. Throughout this work we 
use a Litim-type flat cutoff function \cite{Litim:2001up}
\begin{align}
  r_k(x) = (1-x) \, \theta(1 - x) \,.
\end{align}
We start with the evaluation of $\Tr\left[G_k R_k\right]$ and write
\begin{align}
 \Tr\left[G_k R_k\right] &= 
  \0{1}{16 \pi^2} \Big[ B_0(\bar\Delta) Q_2(R_kG_k)
  \\[1ex]
  &\qqquad+ B_2(\bar\Delta) Q_1(R_kG_k) \Big] 
  + \mathcal O (\bar R^2) \,,\notag
\end{align}
where
\begin{align}
  Q_n[W] &= \0 1{\Gamma(n)} \, \int \mathrm dx \; x^{n-1} \, W(x) \,,\notag \\[1ex]
  B_n &= \int \mathrm d^dx \, \sqrt{\bar g} \; \tr \, b_n \,,
\end{align}
and $\tr \, b_n(\bar\Delta_s) \propto \bar R^{n/2}$ are the  heat kernel coefficients.
We parameterise all graviton and ghost modes from \eqref{eq:graviton-modes} and \eqref{eq:ghost-modes}
with the function
\begin{align}
  W(x) = \0{r_k(x)}{[x+r_k(x)+a \, \lambda]^b} \,,
\end{align}
with constants $a$ and $b$.
This results in
\begin{align}
  Q_2[W] = \0 16 \0 {k^{6-2b}}{(1 + a \lambda)^b} \,,
  \qquad
  Q_1[W] = \0 12 \0 {k^{4-2b}}{(1 + a \lambda)^b} \,.
\end{align}
Furthermore on the sphere we have
\begin{align}
  B_0(\bar\Delta_s) = \sqrt{ \bar g } \;\tr \, b_0 \,, 
  \qqquad
  B_2(\bar\Delta_s) = \sqrt{ \bar g } \;\tr \, b_2 \,, 
\end{align}
where the trace coefficients are given in \autoref{tab:coeff}.
\begin{table}
  \caption{Heat kernel coefficients for transverse traceless tensors (TTT), transverse vectors (VT)
  and scalars (S) on $S^4$.}
  \label{tab:coeff}
 \setlength{\tabcolsep}{7pt}
  \begin{center}
    \begin{tabular}{  c  c  c  c  }
      \hline
      & TTT & VT & S \\
      \hline
      $\tr \, b_0$ & 5 & 3 & 1 \\[1ex]
      $\tr \, b_2$ & $-\0 56 R$  & $\0 14 R$  & $\0 16 R$  \\[1ex]
      \hline
    \end{tabular}
  \end{center}
\end{table}
By specialising the coefficients $a$ and $b$ in the general expression above and evaluating the sum over
all spin modes, we find
\begin{align} \label{eq:sol-Tr-GR}
 \Tr\left[G_k R_k\right] ={}&  \0{ \sqrt {\bar g} }{ 16 \pi^2 }
  \bigg\lbrace
    \bigg[
    \0 56 \0{1}{1 - 2 \lambda}
    + \0 16 \0{1}{1 - \0 43 \lambda}
    - \0 23
    \bigg] k^4 \bar R^0 \notag \\[1ex]
    &+ \bigg[
      - \0 {5}{9} \0{1}{(1 - 2 \lambda)^2}
      - \0 {5}{12} \0{1}{1 - 2 \lambda}
    \notag \\[1ex]
    &\qquad+ \0 {1}{12} \0{1}{1 - \0 43 \lambda}
      - \0 {7}{18}
    \bigg] k^2 \bar R^1 
  \bigg\rbrace
    + \mathcal O (\bar R^2) \,.
\end{align}

We turn now to the evaluation of $\Tr\left[F_{\RG}\right]$,
which is evaluated in similar fashion. We write
\begin{align}
 \Tr\left[F_{\RG}\right] &= \0{1}{16\pi^2} \, F_\RG (-\partial_\tau)
  \Big[ B_0(\bar\Delta_s) \, \tau^{-2} 
  \\[1ex]
  &\qqqquad + B_2(\bar\Delta_s) \, \tau^{-1} \Big] \bigg|_{\tau=0} 
    + \mathcal O(\bar R^2) \,,\notag
\end{align}
where we used
\begin{align}
  \Tr f(\bar \Delta) = f(-\partial_\tau ) 
  \Tr\, e^{-\tau \bar \Delta}\bigg|_{\tau=0} \,.
\end{align}
In particular, by utilising the identities
\begin{align}
  \0 1{\tau^2} &= \int_0^\infty \mathrm dx \; x \, e^{-\tau x} \bigg|_{\tau=0} \,,
  \notag\\[1ex]
  \0 1\tau &= \int_0^\infty \mathrm dx \; e^{-\tau x} \bigg|_{\tau=0} \,,
\end{align}
one evaluates the action of the differential operators and finds
\begin{align}
  F_\RG(-\partial_\tau) \tau^{-2} &= \int_0^\infty \!\mathrm dx \; x \, F_\RG(x)
  = - \0 13 \0{k^{6-2b}}{(1+a\lambda)^b}  \,,\notag\\[1ex]
  F_\RG(-\partial_\tau) \tau^{-1} &= \int_0^\infty \!\mathrm dx \; F_\RG(x)
  = - \0 12 \0{k^{4-2b}}{(1+a\lambda)^b}\,,
\end{align}
where the last equality in each line holds for the general function
\begin{align}
  F_\RG(x) &= \int^{x}_0 \mathrm dy \; \0{r'_k(y)}{(y+r_k(y)+a \, \lambda \, k^2)^b} \notag \\[1ex]
  &=  -\0{x}{(1 + a \lambda)^b } \,,\quad \text{for } x \leq k^2 \,,
\end{align}
using the flat Litim-type cutoff as before.
Specialising the coefficients $a$ and $b$ for the various spin modes and summing the contributions
we get
\begin{align}\label{eq:sol-Tr-FRG}
 \Tr\left[F_{\RG}\right] ={}& \0{ \sqrt {\bar g} }{ 16 \pi^2 }
  \bigg\lbrace
    \bigg[
    - \0 53 \0 {1}{1 - 2 \lambda}
    - \0 13 \0 {1}{1 - \0 43 \lambda}
    + \0 43
    \bigg] k^4  \notag\\[1ex]
    &+ \bigg[
        \0 {5}{12} \0{1}{1 - 2 \lambda}
      - \0 {1}{12} \0{1}{1 - \0 43 \lambda}
      + \0 {5}{24}
    \bigg] k^2 \bar R^1
  \bigg\rbrace
  \notag\\[1ex]
    &+ \mathcal O (\bar R^2) \,.
\end{align}
The evaluated traces \eqref{eq:sol-Tr-GR} and \eqref{eq:sol-Tr-FRG}
allow us to compute the corrections from the Nielsen identity \eqref{eq:I-slim}.
The result is displayed in \eqref{eq:sol-I} where also the contribution from the
minimally coupled scalars is shown. The computation of the latter is 
the same as the scalar graviton modes.

Last we evaluate $\Tr\left[G_k \partial_t R_k\right]$.
which is done in straight analogy to the previous traces.
The heat-kernel functionals are now parameterised with 
\begin{align}
  \tilde W(x) = \0{\partial_t\, k^2 \,r_k(x)}{[x+r_k(x)+a \, \lambda]^b} \,.
\end{align}
With $\partial_t \,k^2 \,r_k(x) = 2 \, k^2 \,r_k(x)$ for the flat cutoff, we find
\begin{align}
  Q_2[\tilde W] = \0 {k^{6-2b}}{(1 + a \lambda)^b} \,,
  \qquad
  Q_1[\tilde W] = \0 {2 \, k^{4-2b}}{(1 + a \lambda)^b}  \,.
\end{align}
where we have suppressed wave-function renormalisations and anomalous dimensions
for readability.
The result of summing over all spin modes is displayed in \eqref{eq:background-flow}.

\bibliography{ScalarGravitylin}
\end{document}